\RequirePackage{fix-cm}
\documentclass[smallextended, natbib]{svjour3}  
\smartqed
\usepackage{url}
\usepackage{graphicx}
\usepackage[caption=false]{subfig}
\usepackage{xspace}
\usepackage{enumitem}
\usepackage{float}
\usepackage[most]{tcolorbox}
\usepackage[title]{appendix}
\usepackage{listings}
\usepackage{color}
\usepackage{natbib}
\usepackage{hyperref}
\usepackage{multirow}
\usepackage{makecell}
\usepackage{xcolor,colortbl}
\usepackage{stfloats}

\definecolor{dkgreen}{rgb}{0,0.4,0}
\definecolor{gray}{rgb}{0.5,0.5,0.5}
\definecolor{mauve}{rgb}{0.58,0,0.82}
\definecolor{rightgreen}{rgb}{0,128,96}

\newcommand\topspace{\rule{0pt}{2.6ex}}            
\newcommand\Tstrut{\rule{0pt}{2.6ex}}    
\newcommand\bottomspace{\rule[-1.5ex]{0pt}{0pt}}  
\newcommand\Bstrut{\rule[-0.9ex]{0pt}{0pt}} 

\newcommand\dataset{\texttt{{JEMMA}}\xspace}
\newcommand\mlfcode{\texttt{{ML4Code}}\xspace}
\renewcommand\cite[1]{\citep{#1}}
\renewcommand\etal{et al.\xspace}    

\usepackage{soul}

\begin{document}

\title{JEMMA: An Extensible Java Dataset for ML4Code Applications}

\author{Anjan Karmakar \and
        Miltiadis Allamanis \and
        Romain Robbes
}

\authorrunning{Karmakar et al.} %
\institute{A. Karmakar \at
              \textit{Free University of Bozen-Bolzano, Italy} \\
              \email{akarmakar@unibz.it}           
          \and
          M. Allamanis \at 
              \textit{Microsoft Research, UK (currently at Google Research)} \\
              \email{miltiadis.allamanis@microsoft.com}
          \and
          R. Robbes \at
              \textit{Free University of Bozen-Bolzano, Italy} \\
              \email{rrobbes@unibz.it} 
}

\maketitle
\begin{abstract}
Machine Learning for Source Code (\mlfcode) is an active research field in which extensive experimentation is needed to discover how to best use source code's richly structured information. With this in mind, we introduce \dataset: An Extensible Java Dataset for \mlfcode Applications, which is a large-scale, diverse, and high-quality dataset targeted at \mlfcode.  %

Our goal with \dataset is to lower the barrier to entry in \mlfcode by providing the building blocks to experiment with source code models and tasks. \dataset comes with a considerable amount of pre-processed information such as metadata, representations (e.g., code tokens, ASTs, graphs), and several properties (e.g., metrics, static analysis results) for 50,000 Java projects from the \texttt{50K-C} dataset, with over 1.2 million classes and over 8 million methods.  

\dataset is also extensible allowing users to add new properties and representations to the dataset, and evaluate tasks on them. Thus, \dataset becomes a {workbench} that researchers can use to experiment with novel representations and tasks operating on source code. 

To demonstrate the utility of the dataset, we also report results from two empirical studies on our data, ultimately showing that significant work lies ahead in the design of context-aware source code models that can reason over a broader network of source code entities in a software project---the very task that \dataset is designed to help with.

\end{abstract}

\section{Introduction}
\label{sec:intro}

Software systems are complex networks of interacting entities. This makes them extremely challenging to develop, understand, and modify---despite the constant need to do so. %
In this context, appropriate tool support for source code can make developers faster and more productive. A variety of such tools have been proposed over the years, ranging from Integrated Development Environments (IDEs), testing tools, static analyzers, version control systems, and issue tracking systems, to name a few. %

\paragraph{Machine learning for source code.} In recent years, significant research effort has been undertaken towards developing tools based on machine learning models of source code \cite{allamanis2018survey} to handle several \textit{tasks}.

{A \textit{task}, in the machine learning paradigm, is a type of action that a machine learning model is trained to perform. Code completion is a good example of a task---that a machine learning model can be trained to perform. 
There can be several other types of tasks, such as code summarization, defect prediction, classification and translation tasks, and many more.}

This line of work was developed from the observation that simple statistical models of source code, such as n-gram models, were surprisingly effective for tasks such as code completion \cite{hindle2016naturalness}. Since then such probabilistic models of source code have come a long way. In the present day, large-scale machine learning models of source code, based on the Transformer architecture, e.g. \texttt{CuBERT} \cite{kanade2020learning}, \texttt{PLBART} \cite{ahmad2021unified}, \texttt{CodeBERT} \cite{feng2020codebert}, and \texttt{GraphCodeBERT} \cite{guo2020graphcodebert} have achieved state-of-the-art performance on a number of Software Engineering (SE) tasks such as code generation, code search, code summarization, clone detection, code translation, and code refinement. 

Largely by increasing the capacity of models and training datasets, deep learning based code completion has transitioned from the token level \cite{karampatsis2020big} to completing entire snippets of code \cite{chen2021evaluating}, the latter being now available on IDEs as an extension named \textit{GitHub Copilot}\footnote{\url{https://copilot.github.com}}.

In parallel, other works on modeling of source code have observed that source code has a well-known structure compared to natural language. Source code can be unambiguously parsed into structured representations, such as  Abstract Syntax Trees (ASTs); functions and methods can have control flows and data flows; functions and methods can interact with each other via calls, parameters and return values. Therefore, even though modeling source code as a series of tokens---analogous to words in a sentence or a paragraph---has proven to be effective, another view shows that accounting for the structure of source code to be more effective.

A fair amount of research has addressed this issue in source code modeling, by proposing the incorporation of the inherent structural information of source code. Several works model source code as Abstract Syntax Trees \cite{mou2016convolutional, alon2018code2seq, leclair2019neural}. Allamanis \etal were among the first to model source code snippets as graphs, including a wide variety of structural information, from data flow information, control flow information, lexical usage information, to call information \cite{allamanis2017learning}.

The space of possibilities to model source code is vast, from text to tokens to advanced graphs---although each comes with its own issues and challenges. Thus, while being mindful of how we represent source code with as much information as possible, we also need to make sure that the models trained on such representations are scalable and reliable for a number of source code tasks and corresponding applications.

\paragraph{From snippets to projects.} 
\label{intro:snippets-projects}

An important limitation of the current breed of deep learning models for source code is that the vast majority of the work has so far focused much more on single code snippets, methods, or functions, rather than on the intricate inter-relationships among source code elements, particularly when these relationships cross file boundaries. 

{Since source code is interconnected and interdependent, we argue that reasoning over a single method or function is fundamentally \emph{inadequate} for several kinds of tasks. For instance, defect prediction tasks, e.g. predicting null pointer exceptions, resource leaks, may benefit from reasoning over associated code entities across the project. In fact, }\citet{10.1145/3360588}{ in their study construct a \textit{global} context by connecting associated method entities based on Program Dependence Graph (\texttt{PDG}) and Data Flow Graph (\texttt{DFG}) to achieve state-of-the-art performance on bug prediction.}

{Even for tasks where the need for additional context may not be apparent, we note that most methods which have multiple calls to other callee methods are in fact dependent on supporting context---since the callee methods logically contribute to the overall functionality of the parent method.} 

{Our views are supported by recent studies which show that encoding additional context while training machine learning models of code significantly improves model performance on a number of tasks. For instance, }\citet{tian2022adding}{ find that adding further context from the call hierarchy (i.e., caller and callee context) improves performance on the clone detection task by 8\%. }\citet{li2021context}{ include additional context from caller-callee methods and sibling methods in the same enclosing class, to train a model on the task of method-naming, and improve upon the state-of-the-art F-score by 11.9\%.} \citet{liu2022learning}{ by encoding a broader context at the project-level, including comments, documentation, and nested scopes, improve on the method-naming task further.} \citet{lu2022reacc}{ make use of additional code with lexical-similarity as external context to establish state-of-the-art performance on the \texttt{CodeXGLUE} benchmark }\cite{lu2021codexglue}{ for code completion task.} 

{In this paper, Section 5 provides further evidence that adding contextual information along with the input representations significantly improves the model performance on a method-call completion task, across four state-of-the-art transformer models: \texttt{BERT}, \texttt{CodeBERTa}, \texttt{CodeBERT}, and \texttt{GraphCodeBERT}.}

{The benefits of including a larger context while modeling source code is demonstrated in the studies mentioned above as well as our own. Therefore, from this current stage, we must gradually move towards building context-aware models that can reason over larger neighborhoods of interacting entities.}

{This is not only a paradigm shift but also a clear indication of the potential need for large-scale code datasets from which additional contexts can be constructed and used in training robust and context-aware source code models.}

The major reasons for the lack of such work is that the necessary data is not yet collected, organized, and is missing at scale, or they support just a single task.
The following section highlights the absence of such datasets for code that have the right mix of source code granularity, size, scale, and detail of information to allow researchers to research on models that go beyond single code snippets. 

Datasets that are large focus either on individual code snippets at the method-level, or at best, source files; while other datasets are either too small, or lack significant preprocessing. Choosing good quality data in sufficient quantity, downloading and storing the data, extracting valuable information from the data or simply running tools to preprocess the data and gather additional information, and then building an experimental infrastructure in place, requires a large amount of time and effort---even before a single experiment is run. This is all the more true when this has to be done for source code models, where some of the pre-processing and analysis tools can be extremely time-consuming and resource-intensive at scale. Therefore, in this paper, we contribute such a dataset: \dataset.

\paragraph{\dataset as a dataset.}
\label{par:jemma-as-a-dataset}
\dataset has multiple levels of granularity: from methods, to classes, to packages, and entire projects. It consists of over 8 million Java method snippets along with substantial metadata; pre-processed source code representations---including graph representations that comes with control- and data-flow information; call-graph information for all methods at the project-level; and a variety of additional properties and metrics. 

{We make sure that all the processed data are clean, consistent, as well as comprehensive---using common data validation, filtering, deduplication, and data curation techniques. Corrupted, incomplete, and blank/null values were corrected where possible; valid results were accurately and consistently mapped to source code entities based on data curation principles; some outputs were filtered at source based on an expected range, on expected datatypes, and/or on formatting conventions; while deduplication eliminated redundant or corrupted entries. Furthermore, the availability of supplementary data down to \texttt{AST}-node level, resulting from our extensive processing, ensured comprehensiveness at scale, for millions of source code entities defined within \texttt{JEMMA}. All of which contribute to the overall quality of the data presented.} 

\dataset is built upon the \texttt{50K-C} dataset of compilable Java projects \cite{martins201850k}, and complements it with significant processing, measured in years of compute time. Section \ref{sec:dataset} presents all the components of the \dataset Dataset.

\paragraph{\dataset as a workbench.} 
\label{par:jemma-intro}
\dataset is not a static dataset: we purposefully designed it to be extensible in a variety of ways. Concretely, \dataset comes with a set of tools to: add metrics or labels to source code snippets (e.g., by utilizing static analysis tools); define prediction tasks based on metrics, properties, or the representation themselves; process the code snippets and existing representations to generate new representations of source code; and run supported models on a task.  We describe how to extend the dataset, along with several examples in Section~\ref{sec:usages}. This extensibility is critical, because it transforms \dataset into a workbench with which users can experiment with the design of ML models of code and tasks, while saving a lot of time in pre-processing the data.

{Traditionally, a database workbench is described as a tool that can be used to view, create, and edit tables, indexes, stored procedures, and other database metadata objects. Thereby, if we now extend the same concept to working with our collection of data, the \texttt{JEMMA} \textit{Workbench} is a set of tools and helper functions that helps in several operations such as viewing, creating, retrieving from, and appending to datasets (independent of how they are stored), among many other tasks that do not involve working directly with the datasets.}

\paragraph{Empirical studies with \dataset.} 
\label{par:jemma-intro-2}
In Sections \ref{sec:non-localness} and \ref{sec:OOW-OOV}, we show how \dataset can be used to gain insights via empirical studies. The first is a study on the \textit{non-localness} of software, and how it impacts the performance of models on a variant of the code completion task. This study shows how the data from \dataset can be used to gain insights into how the models perform on code samples, highlighting what performance issues exist and what we can do to address such issues by adding project-wide context (Section~\ref{sec:non-localness}). 

The second is the study of the size of entities that constitute software projects, and how it relates to the context size of popular machine-learning (ML) models. The second study confirms that significant work lies ahead in designing models that efficiently encode large contexts (Section~\ref{sec:OOW-OOV}). 

{While these examples are related to empirical analyses in the sub-field of Machine Learning for Software Engineering, we can envision further uses for \texttt{JEMMA} in empirical studies. For example, empirical studies on fault-prone or misleading method names, or the impact of complexity on other code properties, or the challenges of coupling in large projects, and others.

Finally, in Section~\ref{sec:limitations} we document the limitations of \texttt{JEMMA}, and then conclude with a summary of our work in Section~\ref{sec:conclusion}.

\section{Related Work}
\label{sec:related}

With the gradual evolution of machine learning techniques suitable for processing code---where data plays a central role---a multitude of efforts have been made for collecting and organizing quality data. Such datasets have not only contributed to the development of competent models of source code, but also opened the avenues for empirical analysis of these models. In this section, we outline some of the datasets from both genres.

\subsection{Datasets for machine learning on code}
\label{sec:rel:ml-datasets}
Since machine learning requires considerable amounts of data, multiple datasets have been produced, usually as a means of validating a specific machine learning method, rather than as a principled standalone effort. This has resulted in datasets that contain input data either not far from raw text, or that contain a lossy view of the underlying analyzed software systems. 

\paragraph{Code Datasets.}
\label{subsec:related:ml-datasets}
\citet{allamanis2013mining} collected a set of over 1 billion Java code tokens and {provide the code text per file for training n-gram models.} Later, \citet{karampatsis2020big} extended this with additional datasets for C and Python; and a different extension of the dataset was provided by \citet{alon2018code2seq}. \citet{raychev2016probabilistic} released Py150 and JS150, two datasets of 150,000 Python and Javascript functions parsed into ASTs. {Unfortunately, these datasets are limited to small programs or code snippets only at the method level. In comparison, \texttt{JEMMA} provides code entities in multiple granularities across several representation types---creating a wide range of modelling opportunities.}

{Several datasets focus on specific tasks,} such as the BigCloneBench~\cite{svajlenko2015evaluating} dataset for large-scale clone detection in Java. ManyTypes4Py~\cite{mir2021manytypes4py} is a Python dataset aimed at evaluating type inference in Python, and Devign~\cite{zhou2019devign}, provides labeled code with coarse-grained source code vulnerability detection in mind. {\texttt{JEMMA}, on the other hand, is not task-specific and supports multiple tasks out of the box.} 

{Datasets with specific representations of code have been common.} CoCoGum ~\cite{wang2020cocogum} use class context represented as abstracted UML diagrams, for code summarization, at the file-level. \citet{allamanis2017learning,allamanis2020typilus} extract control, data flow graphs, along with syntax within a single file. {Representation-specific datasets are useful but they limit cross-representational and cross-architectural analyses for tasks. \texttt{JEMMA} supports several representations including raw source code, tokens, ASTs, and graphs, at both method-level and class-level, building up to even coarser granularities.}

Datasets of code from student assignments, programming competitions, and other smaller programs, have also been created. Among them, Google Code Jam\footnote{\url{https://code.google.com/codejam/contests.html}} and POJ-104~\cite{mou2016convolutional} are clone detection tasks (clone detection in this case is formulated as a program classification task). COSET~\cite{wang2019coset}, and CodeNet~\cite{puri2021project} also feature smaller programs, but complement them with additional metrics and labels. {Although these datasets have many desirable properties, they do not represent source code used in real-life software systems and thus it is unclear if learning on these datasets can generalize to general-purpose software. Semi-synthetic datasets, such as NAPS}~\cite{zavershynskyi2018naps}, {also fall into the same category.} {\texttt{JEMMA} balances the preceding concerns by building upon organic projects coming from a diverse set of domains (e.g., games, websites, standalone applications, etc) and of development standards (ranging from student projects to industry-grade open-source projects) which add a healthy factor of generalization for source code modeling.} 

\paragraph{Code Datasets with Natural Language.}
Natural language presents an interesting, yet separate, modality from source code and is central to the NLP task of semantic parsing (i.e., text-to-code). A few datasets have focused on this:  CodeNN~ \cite{iyer2016summarizing}, CoNaLa ~\cite{yin2018learning}, and StaQC ~\cite{yao2018staqc}. Datasets such as NL2Bash~\cite{lin2018nl2bash} provide data for semantic parsing from natural language to Bash commands, while Spider~\cite{yu2018spider} is a dataset for the text-to-SQL semantic parsing task. Finally, \citet{barone2017parallel},
CodeSearchNet~\cite{husain2019codesearchnet}, and \citet{leclair2019recommendations} pair natural language documentation comments with code, targeting code search and code summarization applications. 

All these datasets provide dumps of source code snippets per file, and while it is possible to parse the code text and perform some intra-procedural analyses for the few file-level datasets, information about external dependencies is commonly lost rendering it impossible to extract accurate semantic data. {\texttt{JEMMA} successfully mitigates such issues by providing a dataset of inter-related code entities across granularities, along with comprehensive intra- and inter-procedural relationship information coming from data-flow, control-flow, call graphs, etc. }

\paragraph{Code Datasets with Higher-level Representations}
\label{sec:rel:larger-context}
While the above datasets focus on code snippets or files, some work has extracted datasets aiming for representations that capture information beyond a single file. {However, commonly these datasets opt for an application-specific representation that loses information that could be useful for other research.} For example, \citet{defreez2018path} extract path embeddings over functions in Linux. LambdaNet~\cite{wei2020lambdanet} extracts type dependency graphs in TypeScript code but removes most code text information; their dataset is also limited to 300 projects, which range from 500 to 10,000 lines of code. The dataset by Bansal \etal was also refined and used in a source code summarization approach that defined a \textit{project-level} encoder, that considers functions in up to 10 source code files in a project \cite{bansal2021project}. {The scale and breadth of information present in \texttt{JEMMA} keeps necessary code information intact, be it across procedures or within procedures, and across files at the project-level.}

\paragraph{Code Datasets with Changes}
\label{par:code-datasets-with-changes}
Given the importance of software maintenance in the development lifecycle, {a few datasets have focused on edit operations in code. The goal of these datasets is to foster research in Neural Program Repair.} ManySStubs4J~\cite{karampatsis2020often} and Bugs2Fix~\cite{tufano2019empirical} both fall in this category: they are corpora of \textit{small} bug fixes extracted from GitHub commits. {These datasets often focus on local changes (e.g., diffs) and ignore the broader context.}

\noindent
{Although our dataset does not come with changes, it provides increased modeling opportunities for users as it comes with inter-procedural relationship information among code entities for all 50K projects. Neural Program Repair workflows could benefit at the dataset creation stage by leveraging bug-related properties in our dataset.}

\subsection{Datasets for empirical studies}
\label{sec:rel:empirical-datasets}

Several corpora of complete software systems have been built with the primary goal to conduct traditional empirical studies, without direct considerations necessary for machine learning research.

\paragraph{The Qualitas Corpus and its descendants.} The Qualitas corpus \cite{tempero2010qualitas}, is an influential corpus of 111 large-scale Java systems that was used for a large number of empirical studies of Java and the characteristics of the systems implemented in it. While this dataset was source code only, it was post-processed in various ways, producing several derived datasets. The Qualitas.class corpus, \cite{terra2013qualitas} is a version of the Qualitas corpus that is also compiled in \texttt{.class} files. The QUAATLAS corpus ~\cite{de2013multi}, is a post-processed version of the Qualitas corpus that allows better support for API usage analysis. XCorpus \cite{dietrich2017xcorpus}, is a subset of the Qualitas corpus (70 programs) complemented by 6 additional programs, that can all be automatically executed via test cases (natural, or generated).

\paragraph{Java Datasets.}  \citet{inproceedings_lammel} gathered a dataset of Java software from Sourceforge, that had 1,000 projects that were parsed into ASTs. The BOA dataset and infrastructure, by \citet{dyer2013boa}, provides an API for pre-processing software repositories, such as providing and analyzing ASTs, for 32,000 Java projects. The 50K-C dataset of \citet{martins201850k} contains 50,000 Java projects that were selected because they could be automatically compiled. A follow-up effort is the Normalized Java Resource~\cite{palsberg2018njr} (NJR). A first release, NJR-1, provides 293 programs on which 12 different static analyzers can run \cite{utture_akshay_2020_4839913}, but has a stated goal of gathering 100,000 \emph{runnable} Java projects, but is still a work in progress.

\paragraph{Other datasets.} 
\label{par:other-datasets}

\citet{spinellis2017repository} released a dataset that contains the entire history of Unix as a single Git repository. \citet{10.1145/3196398.3196460} {present a graph-based dataset of commit history of real-world android apps.} The entire Maven software ecosystem was released as a dataset with higher-level metrics, such as changes and dependencies \cite{raemaekers2013maven}. {The Maven Dependency Graph by} \citet{10.1109/MSR.2019.00060} { provides a snapshot of the Maven Central as a graph, modeling all its dependencies.} Fine-GRAPE is a dataset of fine-grained API usage across the Maven software ecosystem \cite{sawant2017fine}. Finally, both Software Heritage \cite{pietri2019software} and World of Code \cite{ma2021world} are very large-scale efforts that aim to gather the entirety of open-source software as complete and up-to-date datasets. The main goal of World of Code is to enable analytics, while the main goal of Software Heritage is preservation (although it also supports analytics). {The Perceval tool }\cite{inproceedingsperceval}{ also promises automatic and incremental data gathering from almost any tool related to contributing to open source development, among other sources.} {We find that although there are similarities between \texttt{JEMMA} and Perceval, the differences highlight that both tools can complement each other well. Perceval can be used to fetch raw project data from a wide variety of data sources. On the other hand \texttt{JEMMA} focuses on source code, and can be used to take care of the analysis of data, pre-processing of data, task definition, and training of models out of the box.}

\subsection{The 50K-C Dataset}
\label{sec:rel:50k-c}

Having surveyed the landscape of existing datasets, we conclude that most machine learning datasets focus on small-scale entities such as functions, methods, or single classes. The ones that offer higher-level representations are specific and too small in scale. The corpora of systems used for empirical studies provide a better starting point, as they can be pre-processed to extract additional information. Of the existing datasets, the most suitable option that is large enough and that allows the most pre-processing is the \texttt{50K-C} dataset of 50,000 compilable projects. 

Since \dataset builds upon \texttt{50K-C}, we provide detailed background information on it in this section. The \texttt{50K-C} dataset is a collection of 50,000 compilable Java projects, with a total of almost 1.2m Java class files, its compiled bytecode, dependent jar files, and build scripts. 
It is divided into three subsets: 
\begin{itemize}
    \item[$\circ$] \texttt{projects}: It contains the 50,000 java projects, as zipped files. The projects are organized into 115 subfolders each with about 435 projects. %
    
    \item[$\circ$] \texttt{jars}: It contains the 5,362 external jar dependencies, which are required for successful project builds. This is important as missing dependencies is the common cause of failing to compile code at scale. %
    
    \item[$\circ$] \texttt{build\_results}: It contains the build outputs for the 50,000 projects, including compiled bytecode, build metadata, and original build scripts. In addition to the above data, a mapping between each project and its GitHub \texttt{URL} is also provided. The bytecode is readily available for a variety of tasks, such as running static analysis tools, or, if the projects can also be executed, as input for testing, and dynamic analysis tools. %
\end{itemize}

Beyond the size of the dataset, the fact that the projects are compilable is the main reason we chose to build upon \texttt{50K-C}. The extensive pre-processing that we perform on top of \texttt{50K-C} requires the use of static analysis tools, to do things such as call graph extraction, and to extract valuable metrics about the systems. Since the vast majority of static analysis tools operate on bytecode, \texttt{50K-C} was the most suitable option that combines both scale and the ability to automate the analysis at such scale.
 
\paragraph{Selection Criteria.}
\label{par:sel-criteria}
The dataset authors downloaded close to 500k Java projects, attempted to compile all of them, and selected 50k projects among the ones that could be compiled. Two filters were applied: projects that were Android applications were excluded, and projects that were clones were also excluded---using the DéjàVu clone repository \cite{lopes2017dejavu}, and the Sourcerer CC tool \cite{sajnani2016sourcerercc}. We find that the projects have a diverse set of domains (e.g., games, websites, standalone apps, etc), and development levels (ranging from student projects to industry-grade open-source projects).  

The dataset consists of both large-scale projects with as many as 5k classes, and smaller projects with as low as 5 classes. %
While the larger projects are good representatives of real-world projects, the smaller projects are valuable too, since machine learning models of code still need to make significant headway in code understanding which necessitates reasoning on projects across all sizes.

\section{The \dataset Dataset}
\label{sec:dataset}

\label{sec:dataset:overview}
Our goal with the \dataset project is to provide the research community with a large, comprehensive, and extensible dataset for Java that can be used to advance the field of source code modeling. The \dataset datasets consist of a large collection of code samples in varying granularities, with wide-ranging and diverse metadata, a range of supported source code representations, and several properties. In addition, it also includes source code information related to code structure, data-flow, control-flow, caller-callee relationships etc. 

For every project in the \dataset Dataset, we gather data at the project-level, and provide information on all the packages and classes. Furthermore, for every class, we parse and provide data on all the methods---including respective metadata, several representations and properties. The detail of data provided for every method entity is comprehensive, with data at the level of \texttt{AST} with data-flow, control-flow, lexical-usage, and call-graph edges among others. In addition to necessary information, such as line numbers and position numbers of code tokens, supplementary information such as token types, node types, etc, are also provided. More details are presented in the following sections.

{\texttt{JEMMA} also comes equipped with \textit{Workbench} tools that allow users to perform a variety of tasks out of the box, such as: transforming code samples into intermediate source code representations, making tailored selections of entities to define tasks and forming custom datasets, or to run supported models, among others (Section~}\ref{sec:usages}{provides more details).}

\paragraph{Statistics.} The original \texttt{50K-C} dataset contains a total of 50,000 projects. It has 85 projects with over 500 classes (with a maximum of 5549 classes in a project), 1264 projects with 101--500 classes, 2751 projects with 51--100 classes, 10693 projects with 21--50 classes, 14322 projects with 11--20 classes, and 20885 projects with 10 or fewer classes (with a minimum of 5 classes per project). We have collected metadata for all of these projects. Overall, the data consists of 1.2 million Java classes, which define over 8 million unique method entities.

\paragraph{Granularity.}
\label{par:granularity}
{\texttt{JEMMA} supports multiple granularities. We have processed and catalogued data starting from the project-level descending to smaller entities, which means a spectrum of granularites of code can be accessed.} 

{However, since a method is the most basic unit of behavior in an object-oriented program, and it is a self-contained program segment that carries out some specific, well-defined logical task, we collect all the properties at the (\textit{finer}) method level. For instance, method entities can be sampled from the datasets and used independently to run code-analysis tools. In this sense, methods can be considered as our primary entity-of-interest.}

{Starting from the method-level, larger contextual entities at the class-, package-, and project-level can then be created by building upon the smaller entities. And since most prevalent models of source code accept input samples at the method-level, we do provide all source code properties and representations at this (\textit{finer}) method-level by default; properties and representations of larger entities-of-interest can always be built upon smaller entities within it.}

\paragraph{Compilability.} Successful compilation ensures that the source code snippets from the projects have been type-checked and parsed successfully, and are valid Java code. Having full-scale compilable projects gives us the assurance that the source code is complete and self-contained; and thus, all the inter-relationships among code entities can be captured and studied. Additionally, static and dynamic analysis tools can be run to generate information for new code tasks.

Some of the tools that we use to post-process the data require the ability to compile the code, rather than just analyzing compiled code. These tools insert themselves in the compilation process (for instance, Infer~\cite{calcagno2015moving}). %
Therefore, we also have to be able to compile the code on demand. Practically, we found that recompilation was not 100\% replicable. Of the 50,000 projects, we were able to compile about 92\% of the projects; a failed compilation is usually linked to a missing or inaccessible dependency. Nevertheless, 100\% the project entities were successfully processed and catalogued, along with their corresponding properties and representations.

\paragraph{Runtime considerations.} 
\label{par:runtime-considerations}
The analyses and post-processing that we apply to the projects is very computationally intensive for some of the tools. For instance, the analyses run just by a single tool---the Infer static analyzer \cite{calcagno2015moving}---can take on the order of half an hour for a single medium-sized project. {Analyzing 50,000 projects with a number of tools and then post-processing the outputs is thus both time-consuming and resource-intensive.} 

{Since projects can vary significantly in size, depending on whether it's a small project or a large one the processing times can be very different.} {For smaller projects, with 1-20 classes, on average it takes 20-30 minutes of processing time overall. For medium-size projects, with 21-50 classes, it takes over an hour. For large projects, with 51-100 classes, it takes a few hours. For the rest of the projects with more than 100 classes, it can take just 4-5 hours or as much as a couple of days depending on the size of the project.}

{We have gone ahead and done most of the necessary pre-/post- processing and only a small portion (about 6-9\% of the data at the time of writing) is still processing and it will be made available shortly.}

\begin{figure}
  \includegraphics[width=120mm]{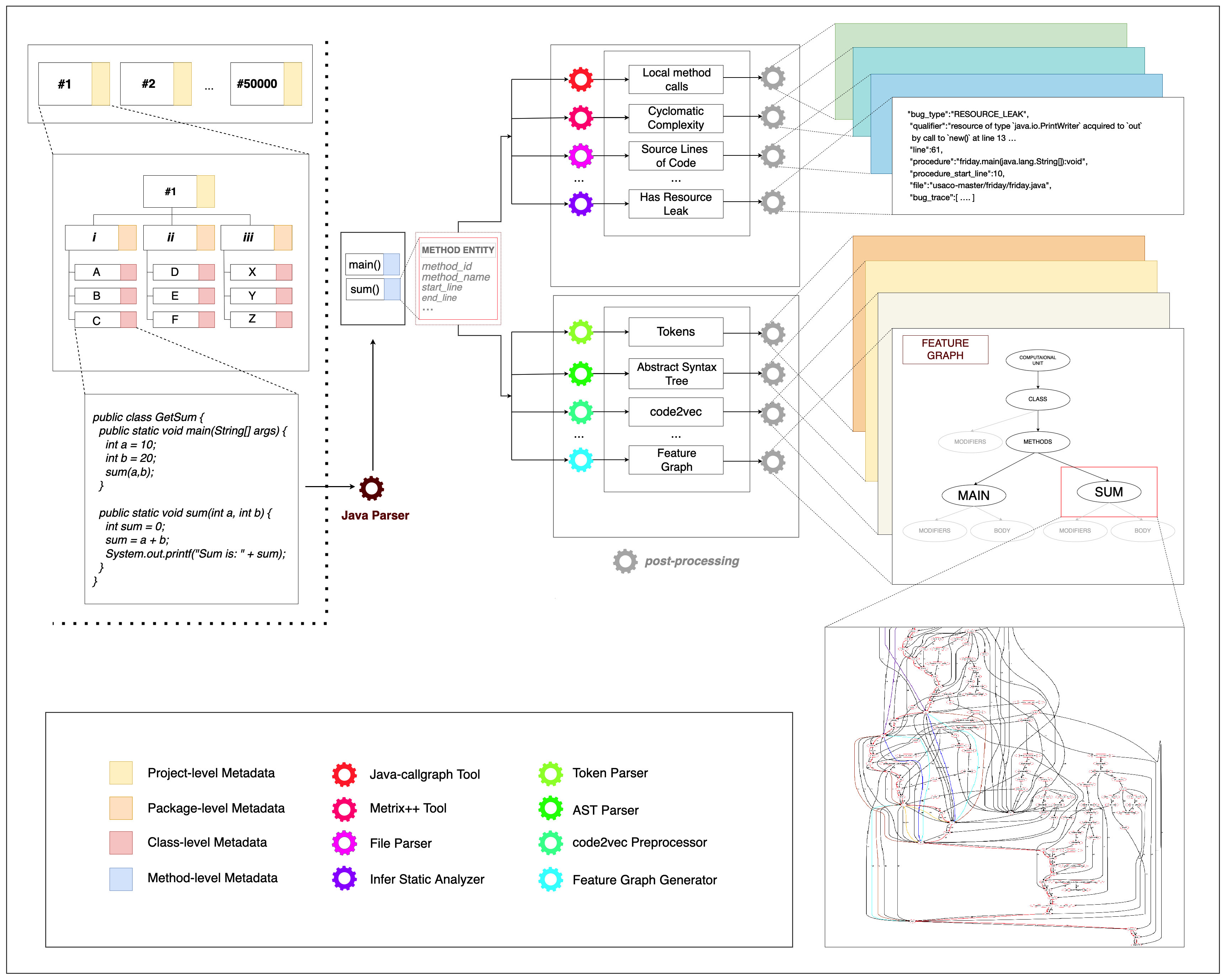}
\caption{Overview of data-level contributions}
\label{fig:overview_a}       
\end{figure}

\paragraph{Storage and sharing considerations.} 
\label{par:storage} 
The amount of data produced is considerable. To maximize accessibility, we provide it as a set of Comma Separated Values (CSV) files, so that users can choose and download the data that they need. Note that only the metadata and the original source data are absolutely necessary; other data can be downloaded on a per-need basis. The \dataset {\textit{Workbench}} allows the recomputation of the other properties, if, for some properties, it is more efficient to recompute them than to download them. The data is uploaded on Zenodo; due to its size, it is provided as multiple artifacts. Table~\ref{tab:datasets} presents the components of the dataset, along with their DOIs (links to the download page), and sizes.

\paragraph{Interacting with the data.} Most of the data from the \dataset are organized in Comma-Separated-Values (CSV) files, consequently basic analyses can be run with tools such as \textit{csvstat}. Furthermore, our \textit{Workbench} \texttt{APIs} can be used to gather extensive statistics of the projects, classes, methods, bytecode, and data and control-flow information.

\noindent
\paragraph{} The \dataset datasets are grouped into three major parts: data at the metadata level (Section~\ref{sec:dataset:metadata}), data at the property level (Section~\ref{sec:dataset:properties}), and at the representation level (Section~\ref{sec:dataset:representations}). In addition, we also provide project-wide callgraph information for the 50,000 projects, uniquely identifying and associating source and destination nodes in the callgraph with the help of the metadata defined by \dataset (Section~\ref{sec:dataset:callgraphs}). This allows for accessing project-wide data on the whole, for different granularities of code entities. 

Fig.~\ref{fig:overview_a} gives a glimpse of the extent and detail of the data contribution made by \dataset. The top-left corner represents the raw data from \texttt{50K-C}, which we catalog by adding \texttt{UUIDs} (symbolized by colored squares). The rest of the figures depicts the additional pre-/post-processing we performed: the colored gears represent external tools that we run to collect additional data (properties and representations), while the grey gears represent further post-processing that we perform on the tool outputs to integrate it in our dataset.

\subsection{{\texttt{JEMMA}: \texttt{Metadata}}}
\label{sec:dataset:metadata}
In this section we present the metadata for the \dataset datasets. The metadata is made available in \textit{CSV} ({comma-separated values}) files. This allows for easy processing, even with simple command-line tools. The metadata is organized in four parts, from the largest units of interest to the smallest: projects, packages, classes, and methods. The units of interest can then be inter-related systematically. The metadata serves two major purposes:

\begin{enumerate}
    \item Uniquely identify a source code entity with an \texttt{UUID}.
    \item Gather basic and often-used information on each source code entity.
\end{enumerate}

Taken together, these two purposes allow users to extend the dataset with additional properties. The \texttt{UUID} allows us to uniquely identify an entity in the dataset, and the supplementary metadata helps disambiguate entities (file paths, parent relationships, location information in the file, etc). In Section~\ref{sec:usages} we show how this metadata can be used to add an additional property to source code entities.

Since the metadata formalizes the organization of the data, and  establishes the relationship between projects, packages, classes, and methods, \dataset users can leverage it to construct custom data queries and make selections from the large collection of data at different granularities. 

\paragraph{Projects.}
For the project-level metadata, we provide a single CSV file that lists all the projects in the \texttt{50K-C} dataset along with their corresponding metadata---\textit{project\_id}, \textit{project\_path}, \textit{project\_name}. The \texttt{UUID} is referenced by the entities contained in the project. The project path is relative to the root directory of the \texttt{50K-C} dataset\footnote{\url{http://mondego.ics.uci.edu/projects/jbf/downloads/50K-C_projects.tgz}}, and can be used to access the raw source code of the project. 

\paragraph{Packages.}
For the package-level metadata, a single CSV file lists all the packages present in the projects. The metadata comprises of the \texttt{UUID} of the parent project as \textit{project\_id}, the \texttt{UUID} assigned to the package as \textit{package\_id}, the relative path of the package as \textit{package\_path}, and the name of the package directory as \textit{package\_name}. 

\paragraph{Classes.}
For the class-level metadata, we provide a single {CSV} file that lists all the classes in the \texttt{50K-C} dataset along with their corresponding metadata: \textit{project\_id}, \textit{package\_id}, \textit{class\_id}, \textit{class\_path}, \textit{class\_name}.  Similarly to the projects, the class path is a relative path starting from the \texttt{50K-C} dataset's root directory, that allows to access the raw source code of the class.

\paragraph{Methods.}
At the method-level, the metadata is more extensive. Just having the name of a method might not be enough to disambiguate methods. Thus, the metadata is a {CSV} file lists all the methods in the \texttt{50K-C} dataset along with their corresponding metadata: \textit{project\_id}, \textit{package\_id}, \textit{class\_id}, \textit{method\_id}, \textit{method\_path}, \textit{method\_name}, \textit{start\_line}, \textit{end\_line}, \textit{method\_signature}. 

\subsection{\dataset: \texttt{Properties}}
\label{sec:dataset:properties}

\dataset leverages the \texttt{UUIDs} assigned to projects, classes, and methods as a way to attach additional properties to these entities. Thus, a property can be an arbitrary value that is associated to an entity, such as a metric. Even though we have gathered several properties associated with code entities, it should be noted that a particular property may not be available or may not apply for a given code entity. Users can add new properties associated with code entities as contributions to the dataset, where the property should be given a unique name and be stored in the correct location for it to be visible to {\texttt{JEMMA} \textit{Workbench} \texttt{APIs}}. (Section \ref{sec:usages} provides more details).

\noindent
Next, we list the tools used to obtain the properties:
\label{par:tool:motivation}
\begin{itemize}
    \item[$\circ$] The \textit{Infer} static analyser \cite{calcagno2015moving} is a tool that provides advanced abstract interpretation-based analyses for several languages, including Java. Examples of the analyses that Infer can run include an interprocedural analysis to detect possible null pointer dereferences. Infer can also perform additional analyses such as taint analysis, resource leak detection, and estimate the run-time cost of methods. {We chose Infer mainly because it can perform inter-procedural analysis that reasons across procedure boundaries, while being able to scale to large codebases.}

    \item[$\circ$] \textit{Metrix++} is a tool that can compute a variety of basic metrics on source code entities, such as lines of code, code complexity, and several others\footnote{\url{https://metrixplusplus.github.io/metrixplusplus/}}. {We chose Metrix++ since it is suitable for processing large codebases, processing thousands of files per minute; it recognizes various types of entities including classes, interfaces, namespaces, functions, comments; and supports multiple metrics.}

    \item[$\circ$] \textit{PMD} is a static code analysis tool\footnote{\url{https://pmd.github.io}} that can compute a variety of static analysis warnings and metrics, such as the \textit{{npath} complexity} metric, among many, many others. {We used PMD because it is inexpensive while reviewing large codebases; and it is trusted by industry practitioners and researchers. PMD can also be used to identify defects and problems in code entities which can be useful for future works.}
    
    \item[$\circ$] The \textit{java-callgraph}\footnote{\url{https://github.com/gousiosg/java-callgraph}} extractor is a tool for generating call graphs in Java. We used this tool to extract project-wide call graphs, from which callers and callees were identified and linked to their respective \texttt{UUIDs} at the post-processing stage. {The java-callgraph generator tool was used since it was capable of generating both static and dynamic call-graphs suitable for our dataset of compilable code entities.}

\end{itemize}

\noindent
Table~\ref{tab:tools:props} provides the list of properties that are currently defined at the finer method-level granularity in \dataset. It also maps the tools used to obtain the properties. Later, at the end of this section, Table~\ref{tab:datasets} provides links to the datasets for all of the data. 

Other tools that could be run to extend the dataset include static analysis tools, such as FindBugs \cite{hovemeyer2004finding}, SpotBugs, or similar tools such as Error Prone and NullAway.
The warnings and outputs from these tools can serve as metrics for code entities. Bespoke static analyses from Soot or other static analysis research frameworks, or clone detection \cite{cordy2011nicad} tools could be run as well. These properties could be useful to conduct studies similar to the ones from \citet{habib2018many}.

\begin{table}
\caption{{List of tools used and properties obtained.}}
\label{tab:tools:props}
\begin{tabular}{|l|l|}
\hline
Property & Tool used \topspace \bottomspace  \\
\hline

\texttt{[TLOC]} Total Lines of Code              & Metrix++   \topspace \\
\texttt{[SLOC]} Source Lines of Code             & Metrix++   \\
\texttt{[CMPX]} McCabe or Cyclomatic Complexity  & Metrix++   \\
\texttt{[MXIN]} Maximum indent depth of nesting  & Metrix++   \\
\texttt{[NPTH]} Npath Complexity                 & PMD        \\
\texttt{[NMTK]} Number of Code Tokens            & Parser     \\
\texttt{[NMPR]} Number of parameters             & Parser     \\    
\texttt{[NUID]} Number of unique identifiers     & Parser     \\
\texttt{[NMOP]} Number of operators              & Parser     \\
\texttt{[NMLT]} Number of literals               & Parser     \\
\texttt{[NMRT]} Number of return statements      & Parser     \\
\texttt{[NAME]} Name of source code entity       & Parser     \\
\texttt{[NUPC]} Number of unique parent callers  & java-callgraph   \\
\texttt{[NUCC]} Number of unique child callees   & java-callgraph   \\
\texttt{[NMNC]} Number of non-local calls        & java-callgraph   \\
\texttt{[NMLC]} Number of local calls            & java-callgraph   \\
\texttt{[NLDF]} Presence of Null Dereference     & Infer      \\
\texttt{[RSLK]} Presence of Resource Leak        & Infer      \bottomspace \\

\hline
\end{tabular}
\end{table}

\subsection{\dataset: \texttt{Representations}}
\label{sec:dataset:representations}

Machine learning models are trained on a collection of feature vectors derived from the input data. For source code machine learning models the input data can be the raw text of a source code entity. For example, for the Java method shown in Figure~\ref{fig:sample_method_text}, a corresponding source code representation could be its raw tokens as shown in Figure~\ref{fig:sample_method_tokens}. 

\begin{figure}[h]
    \centering
    \subfloat[Java method as original text representation]{
    \frame{\includegraphics[width=95mm, scale=0.8]{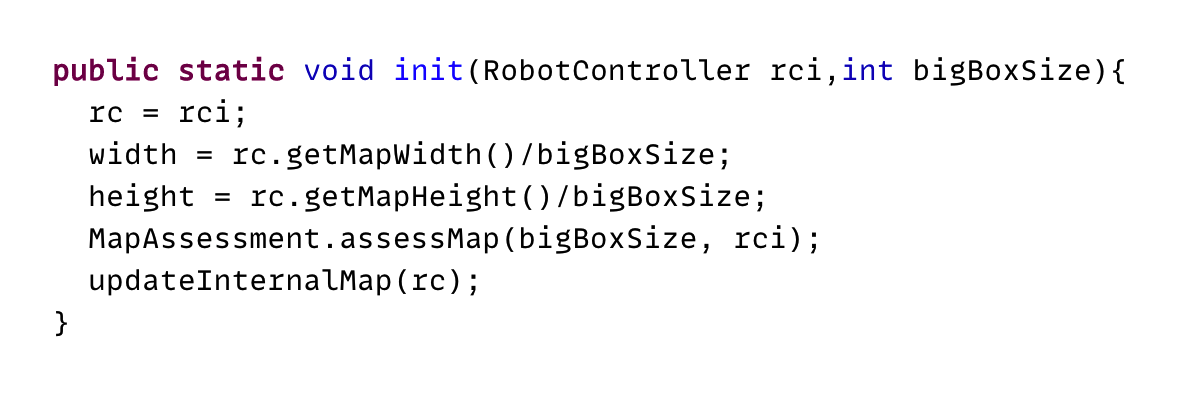}}
    \label{fig:sample_method_text}
    }

    \subfloat[Java method represented as tokens]{
    \frame{\includegraphics[width=95mm, scale=0.8]{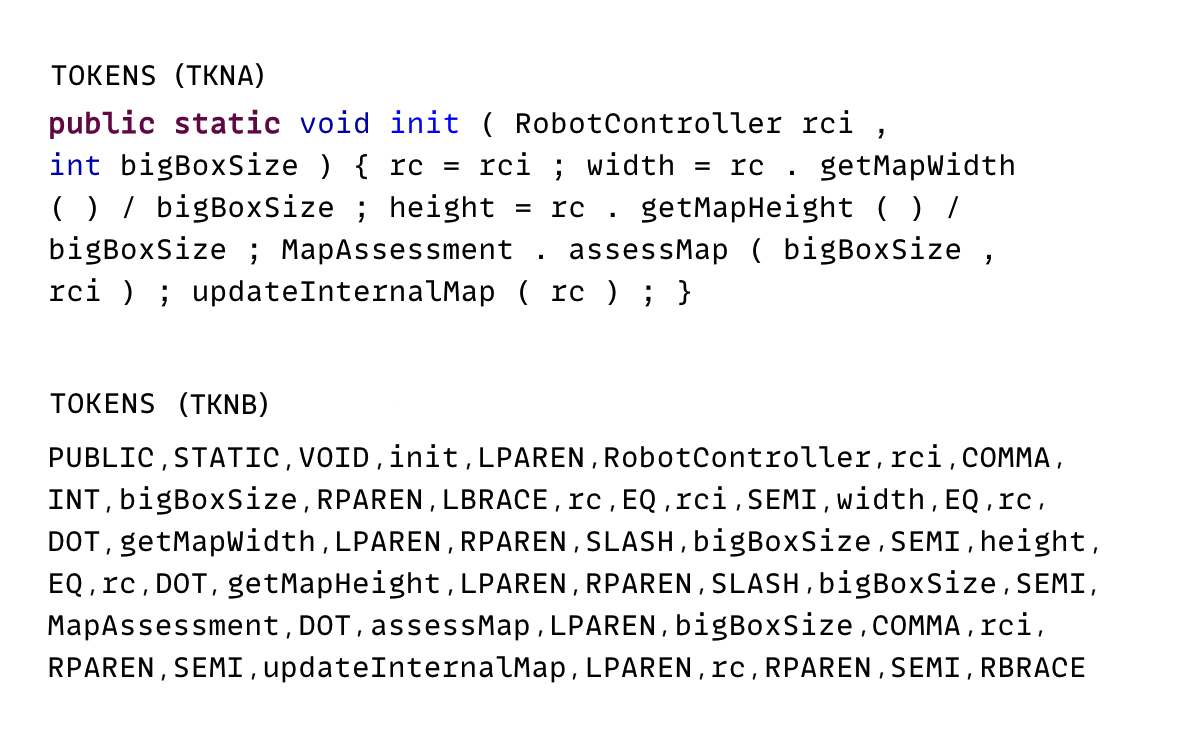}}
    \label{fig:sample_method_tokens}
    }
    \caption{An example of a Java method and two of its possible token representations}
\end{figure}

\noindent
Since source code is highly structured, the design space for representations is vast and diverse. This has been explored to some extent, with approaches that model source code as sequences of tokens or subtokens via RNNs \cite{pradel2018deepbugs}, LSTMs \cite{karampatsis2020big}, or Transformers \cite{feng2020codebert}. Other approaches have leveraged the structure of code either via ASTs \cite{mou2016convolutional} or linearized ASTs \cite{alon2019code2vec}. Yet other approaches use more expressive structures incorporating, for instance, data flow, and use Graph Neural Networks (GNNs) to represent code \cite{allamanis2017learning}. 

Our goal with \dataset is to provide the building blocks to experiment with the design space of representations. Since extracting the relevant information is costly in terms of computational resources, a significant effort went into adding several basic representations at the method level, ranging from the raw source code to the information behind a very complete graph representation. At the representation level, we provide several ready-to-use source code representations (compatible with different models) for over 8 million method snippets. The method level representations that we provide are described in the following subsections.

\subsubsection{Raw text (\texttt{TEXT})} 
First and foremost, the original code for each method is provided by default, with no preprocessing. This allows approaches that need to operate on the raw text to do so directly (e.g., a model that implements its own tokenization). The default method text includes its annotations and also comments within the code snippet, if any. The whitespace for each method text is also preserved intentionally (it can be easily stripped off at any point). The raw text can also be used to re-generate the other representations as needed.

\subsubsection{Tokens (\texttt{TKNA}, \texttt{TKNB})}

Each Java method snippet is tokenized with an \texttt{ANTLR4} grammar \cite{parr2013definitive} and made available to the user preprocessed. The tokenized code includes method annotations, if any, but does not include natural language comments. However, with the entire raw text of method snippets made available by default, users are free to include comments in their custom tokenizations.  

For every method snippet in our dataset, we provide the corresponding string of tokens. In fact, we provide two types of tokenized strings. First, a simple space-separated string of tokens. This representation is meant to be directly passed to an existing ML model that has its own tokenizer, without any further processing. The downside is that some literals that include spaces may be treated as more than one token, or symbols and special characters may be ignored while using certain tokenizers (e.g., natural-language tokenizers). Should this be an issue, users may use the second representation. 

In the second type of tokenized representation the tokens are made available as a comma-separated string with the symbols and operators replaced with suitable string tokens (commas in literals are replaced suitably with \texttt{<LITCOMMA>} tokens). This representation is recommended for users who would tokenize the code themselves, or would want to avoid literals being split into several tokens, or avoid ambiguities with symbols and special characters when using natural-language tokenizers.

\subsubsection{Abstract Syntax Tree (\texttt{ASTS})}

An Abstract Syntax Tree, or AST, is a tree representation of the source code of a computer program that conveys the structure of the source code.

In an AST, nodes can either be terminal nodes (leaf nodes), which are the tokens of the grammar, or non-terminal nodes (internal nodes), representing the higher-level units such as method calls, code blocks, etc. This information is represented for each method as a set of nodes, followed by a set of node pairs representing child edges.

\subsubsection{code2vec (\texttt{C2VC}) and code2seq (\texttt{C2SQ})} 
\label{subsec:c2v:c2s}
The code2vec \cite{alon2019code2vec} and code2seq \cite{alon2018code2seq} representations are derivatives of the AST representation. The goal of these approaches is to linearize ASTs by sampling a fixed number of AST paths (i.e., selecting 2 AST nodes at random and utilizing the connected path between them). 

The difference between the approaches is that code2vec represents each identifier and each path as unique symbols leading to large vocabularies, and consequently Out-Of-Vocabulary (\texttt{OOV}) issues, while code2seq models identifiers and paths as sequences of symbols from smaller vocabularies, which alleviates the same issues. However, the downside is that the code2seq representation is significantly larger. Both kinds of inputs are fed to models that use the attention mechanism to select a set of AST paths that are relevant to the model's training objective (by default, method naming).

{We have generated the code2vec and code2seq representations of every method in the dataset by running the pre-processing scripts, which can serve as a ready-to-use input to the code2vec and code2seq path-attention models.} Furthermore, if mutations to the code snippets are necessary, our \textit{Workbench} tools enable users to easily transform raw code snippets into the corresponding representations using the original code2vec and code2seq preprocessors.

\subsubsection{Feature Graph (\texttt{FTGR})} 
\label{sec:representations:ftgr}

{A Feature Graph is a feature-rich graph representation of a source code entity. It is built on top of the abstract syntax tree, but containing multiple edge types to model syntactic, semantic, lexical, data-flow, and control-flow relationships between nodes.} \cite{allamanis2017learning} 

The Feature Graph representation is comprised of a set of nodes, and then node pairs representing different edge types; the nodes are also presented in a sequence to capture the order of tokens. Specific edge types can be filtered as needed (such as to produce the AST representations, or to reduce the size of the graph \cite{hellendoorn2019global}). The full list of included edges are:

\begin{itemize}
    \item[$\circ$] \texttt{Child} edges encoding the \texttt{AST}.
    \item[$\circ$] \texttt{NextToken} edges, encoding the sequential information of code tokens.
    \item[$\circ$] \texttt{LastRead}, \texttt{LastWrite}, and \texttt{ComputedFrom} edges that link variables together, and provide data flow information.
    \item[$\circ$] \texttt{LastLexicalUse} edges link lexical usage of variables (independent of data flow).
    \item[$\circ$] \texttt{GuardedBy} and \texttt{GuardedByNegation} edges connecting a variable used in a block to conditions of a control flow.
    \item[$\circ$] \texttt{ReturnTo} edges link return tokens to the method declaration.
    \item[$\circ$] \texttt{FormalArgName} edges connect arguments in method calls to the formal parameters.
\end{itemize}

This representation is significantly feature-rich, as it includes, for instance, all the source code tokens and their precise locations in the original source code, the signatures of all the methods called in the class, the source code comments, if any, including a variety of data-flow, control-flow, lexical-usage, and hierarchical edges. Derivative representations such as AST with dataflow information, a subset of the feature graph representation, which corresponds to the data used by models such as \texttt{GraphCodeBERT} can also be produced from the feature graph representations. The feature graph representation is obtained from Andrew Rice's feature graph extraction tool\footnote{\url{https://github.com/acr31/features-javac}}.

\subsection{\texttt{JEMMA: Callgraphs}}
\label{sec:dataset:callgraphs}

Since many relationships among source code entities are not simply hierarchical containment relationships, we also provide a very useful additional data: the project's call graph (\texttt{CG}), in which methods calling each other are explicitly linked. Thanks to our metadata, these method call information 
can then be used to combine representations to create interesting global contexts for large-scale source code models.  

Previous techniques are useful to design representations at the level of methods. However, designing models that reason over larger entities requires more data. Hierarchical relationships can be already inferred from the meta-data. In addition, since software systems are composed of modules that interact with each other, caller-callee relationships are crucial to model systems accurately. For this, we use a Java callgraph extractor tool, to extract project-wide call graphs, from which callers and callees are identified and linked to their respective \texttt{UUIDs} through post-processing (links to external calls are still recorded but we do not assign \texttt{UUIDs} to them). 

Method signatures are used to disambiguate methods with similar names. Note that for polymorphic calls, the call graph provides links to the statically identified type, not to all possible types.  Additional post-processing would be possible to add these links to the call graph. In previous work, we have seen that the use of polymorphism in Java is significant \cite{milojkovic2015polymorphism}, so this would be a useful addition.

\begin{table}
\caption{{\texttt{JEMMA} dataset artifacts, locations, and sizes}}
\label{tab:datasets}
\begin{tabular}{|l|c|r|r|}
\hline
Artifact & DOI & Size \topspace \bottomspace  \\
\hline
\textit{Metadata: Projects}  & \url{https://doi.org/10.5281/zenodo.5807578} & 4.7 MB\Tstrut{}\\
\textit{Metadata: Packages}  & \url{https://doi.org/10.5281/zenodo.5807586} & 42.2 MB  \\
\textit{Metadata: Classes}   & \url{https://doi.org/10.5281/zenodo.5808902} & 269.7 MB  \\
\textit{Metadata: Methods}   & \url{https://doi.org/10.5281/zenodo.5813089} & 2.8 GB\Bstrut{}\\ 
\hline

\textit{Properties:} \texttt{[TLOC]}  & \url{https://doi.org/10.5281/zenodo.5813102} & 335.5 MB\Tstrut{}\\
\textit{Properties:} \texttt{[SLOC]}  & \url{https://doi.org/10.5281/zenodo.5813094} & 335.0 MB  \\

\textit{Properties:} \texttt{[NUID]}  & \url{https://doi.org/10.5281/zenodo.5813028} & 335.6 MB  \\
\textit{Properties:} \texttt{[NTID]}  & \url{https://doi.org/10.5281/zenodo.5813029} & 336.7 MB  \\ 
\textit{Properties:} \texttt{[NMTK]}  & \url{https://doi.org/10.5281/zenodo.5813032} & 342.5 MB  \\
\textit{Properties:} \texttt{[NMRT]}  & \url{https://doi.org/10.5281/zenodo.5813034} & 333.3 MB  \\
\textit{Properties:} \texttt{[NMPR]}  & \url{https://doi.org/10.5281/zenodo.5813053} & 333.3 MB  \\
\textit{Properties:} \texttt{[NMOP]}  & \url{https://doi.org/10.5281/zenodo.5813055} & 334.5 MB  \\ 
\textit{Properties:} \texttt{[NMLT]}  & \url{https://doi.org/10.5281/zenodo.5813054} & 333.4 MB  \\
\textit{Properties:} \texttt{[NAME]}  & \url{https://doi.org/10.5281/zenodo.5813308} & 432.0 MB  \\
\textit{Properties:} \texttt{[MXIN]}  & \url{https://doi.org/10.5281/zenodo.5813081} & 267.0 MB  \\
\textit{Properties:} \texttt{[CMPX]}  & \url{https://doi.org/10.5281/zenodo.5813084} & 267.1 MB\Bstrut{}\\

\textit{Properties:} \texttt{[NUPC]}  & \url{https://doi.org/10.5281/zenodo.7019128} & 333.3 MB  \\
\textit{Properties:} \texttt{[NUCC]}  & \url{https://doi.org/10.5281/zenodo.7019176} & 333.6 MB  \\

\textit{Properties:} \texttt{[NMNC]}  & \url{https://doi.org/10.5281/zenodo.7019960} & 334.0 MB  \\
\textit{Properties:} \texttt{[NMLC]}  & \url{https://doi.org/10.5281/zenodo.7020084} & 333.2 MB  \\

\textit{Properties:} \texttt{[NLDF]}  & \url{https://doi.org/10.5281/zenodo.1096080} & 333.6 MB  \\
\textit{Properties:} \texttt{[RSLK]}  & \url{https://doi.org/10.5281/zenodo.1096082} & 334.0 MB\Bstrut{}\\ 
\hline

\textit{Represent.:} \texttt{(TEXT)} & \url{https://doi.org/10.5281/zenodo.5813705} & 3.8 GB\Tstrut{}\\
\textit{Represent.:} \texttt{(TKNA)} & \url{https://doi.org/10.5281/zenodo.5813717} & 3.3 GB \\
\textit{Represent.:} \texttt{(TKNB)} & \url{https://doi.org/10.5281/zenodo.5813730} & 4.6 GB \\
\textit{Represent.:} \texttt{(ASTS)} & \url{https://doi.org/10.5281/zenodo.5813880} & 4.1 GB \\
\textit{Represent.:} \texttt{(FTGR)} & \url{https://doi.org/10.5281/zenodo.5813933} & 5.2 GB \\ 
\textit{Represent.:} \texttt{(C2VC)} & \url{https://doi.org/10.5281/zenodo.5813993} & 6.1 GB \\
\textit{Represent.:} \texttt{(C2SQ)} & \url{https://doi.org/10.5281/zenodo.5814059} & 10.9 GB\Bstrut{}\\
\hline

\textit{Callgraphs: Projects}  &
\url{https://doi.org/10.5281/zenodo.6758937} & 7.2 GB\Tstrut{} \Bstrut{}\\
\hline 

\end{tabular}
\end{table}

\section{Extending and Using \dataset}
\label{sec:usages}
Table~\ref{tab:datasets} presents the links to the actual datasets with \dataset: meta-data, properties, representations, and callgraphs. These are standalone \texttt{CSV} files that can be used on their own, but to make it easy for users to access and use them in common usage scenarios we have added a \textit{Workbench} component to \dataset.

When building \dataset, we intended it to be large-scale, yet extensible, flexible, and most importantly, easy to use. We have implemented several tools to help with this, and as a result, researchers can readily use \dataset as a \textit{Workbench} to experiment with variants of datasets, models, and tasks while minimizing the processing that is involved.

{The \texttt{JEMMA} \textit{Workbench} tools and implementations, written in Python, are accessible through a set of \texttt{APIs}, which helps developers interface with it when writing machine-learning code and take advantage of several pre-implemented functionalities from viewing, creating, retrieving from, and appending to datasets, defining task labels, generating custom/variant code representations, to training and evaluating supported models on such datasets.}

\noindent
{On a high level, the \textit{Workbench} supports the following types of operations. The next sections describe the usage of the \textit{Workbench} tools to perform some of these operations in more detail.} 
\begin{itemize}
    \item[$\circ$] {\texttt{GET} meta-data, properties, representations, callers/ees, n-hop context}
    \item[$\circ$] {\texttt{ADD} meta-data, properties, representations, callers/ees}
    \item[$\circ$] {\texttt{GEN} (create/adapt) representations} 
    \item[$\circ$] {\texttt{RUN} (train/evaluate) supported models on a task}   
\end{itemize}

\vskip 1mm %
\noindent
{These operations are supported by a set of \texttt{API}s. The \texttt{JEMMA} \textit{Workbench} implementations, along with an exhaustive list of \texttt{API}s, are made available online.\footnote{\url{https://github.com/giganticode/jemma/}} We demonstrate the usage of some of the \texttt{API}s in this section.}

\vskip 1mm %
\noindent
{\textbf{Potential uses for the data.} In the following paragraphs, we describe some of the potential ways in which data can be utilized directly with the help of the Workbench tools.} 

{\textit{Create a multitude of new datasets:} Since \texttt{JEMMA} is built on top of the \texttt{50K-C} Dataset, we have catalogued all of the 50,000 projects and their child code entities, made them uniquely-identifiable, and provided a range of properties associated with them, along information on inter-relationships. Using this information a multitude of datasets can be prepared, not just specifically for \texttt{ML4Code} but also for other purposes, e.g. creating a dataset of projects based on the project size, or creating a dataset of method snippets with complexities based on a criterion, and so on, for a diverse range of use-cases.} 
    
{\textit{Define ML task datasets:} Users can decide on a modelling objective and retrieve the training data from \texttt{JEMMA}. This may sound straightforward but preparing a sound dataset for model training is one of the most important steps, and it is often time-consuming given the amount of data cleaning and transformations involved before model training. The \texttt{JEMMA} \textit{Workbench} tools help users choose from a range of pre-processed source code representations across 8M  samples, and filter them based on a range of properties, and even use the properties as prediction labels. The representations and properties can be used, either singularly, or in combination, to generate thousands of combinations of clean and balanced ML task datasets ready to be trained. Section 4.2.1 demonstrates a similar example.}   
    
{\textit{Retrieve code information:} Within \texttt{JEMMA} we catalogue code at the project, package, class, and method-level. Furthermore, we process source code into feature graphs yielding feature-rich code information at the AST node-level. This enables users to access a diverse range of granularites from coarse file-level to finer node-level information. Not only that, information such as the data-flow, control-flow, etc. between nodes are also available at the node level---providing a remarkable level of detail for code entities. In addition, call-graph links provide information on the inter-procedural relationships across entities within projects. This affords users to access code entities at scale, in different granularities, with detailed and intricate information based on various intra- and inter-procedural relationships.}      

\vskip 1mm %
\noindent
{ \textbf{Potential operations on the retrieved data.} Here we describe potential operations that can be performed on the data once they have been queried.}

{\textit{Working with representations.} Different model architectures process input in different formats. Graph models work with code represented as graphs while some others may process code represented as ASTs. Using the \textit{Workbench} tools users can easily undertake operations such as extensions and abstractions. For example, users can abstract from an existing representation to create sparse representations (e.g. from feature graphs to just data-flow graphs) or add new information to existing representations to make it more feature-rich. Our tools help users create abstracted or extended alternatives of existing representations. In addition, the \textit{Workbench} tools make it extremely easy for users to re-generate representations from scratch when necessary (see Section 4.2.2).}
    
{\textit{Conduct analyses.} Having accessed the data at scale, one of the most obvious things that users can do is to conduct statistical analyses. In addition, users can conduct a multitude of empirical studies to test a range of hypotheses utilizing the aggregated array of information on millions of code entities.}
    
{\textit{Training and evaluating models.} Since \texttt{JEMMA} was prepared with \texttt{ML4Code} in mind, users can easily train/evaluate a number of models, conduct inference, and establish benchmarks for tasks. The diversity of representations facilitates training on several different types of model architectures, from graph-based models, to models that take ASTs as input, and other architectures such as code2seq which reason over a bag of AST-paths. This enables users to model source code in various formats and combinations and extract valuable insights.}

In the next sections, we concretely demonstrate how \dataset can be extended and used, emphasizing on some essential use-cases.

\subsection{Extending \dataset}
\label{sec:extending-jemma}
In Section~\ref{sec:usages:define-property} we describe how \dataset can be extended with a new property; in Section~\ref{sec:usages:define-representation} we describe how it can be extended with a new representation; and in Section~\ref{sec:usages:adding-project} we show new projects can be added to \dataset.

\subsubsection{Adding a new property}
\label{sec:usages:define-property}

The simplest way to extend \dataset is to add a new property. This could be any property of interest that can be computed for a source code entity. Examples include defining a new source code metric, or the result of a static analysis tool indicating the presence (or absence) of a specific source code characteristic. %

To extend \dataset with a new property, the workflow has three main steps: a) accessing a set of source code entities, b) generating associated property values, and c) merging the associated property values to the dataset. \dataset facilitates accessing the correct code input by providing the location and metadata for code entities, and several initial representations (raw text, \texttt{ASTs}, etc.). An associated property could then be obtained either directly (e.g. method name) or by means of a code analysis tool (e.g. cyclomatic complexity).

\begin{figure}[H]
  {\includegraphics[width=110mm]{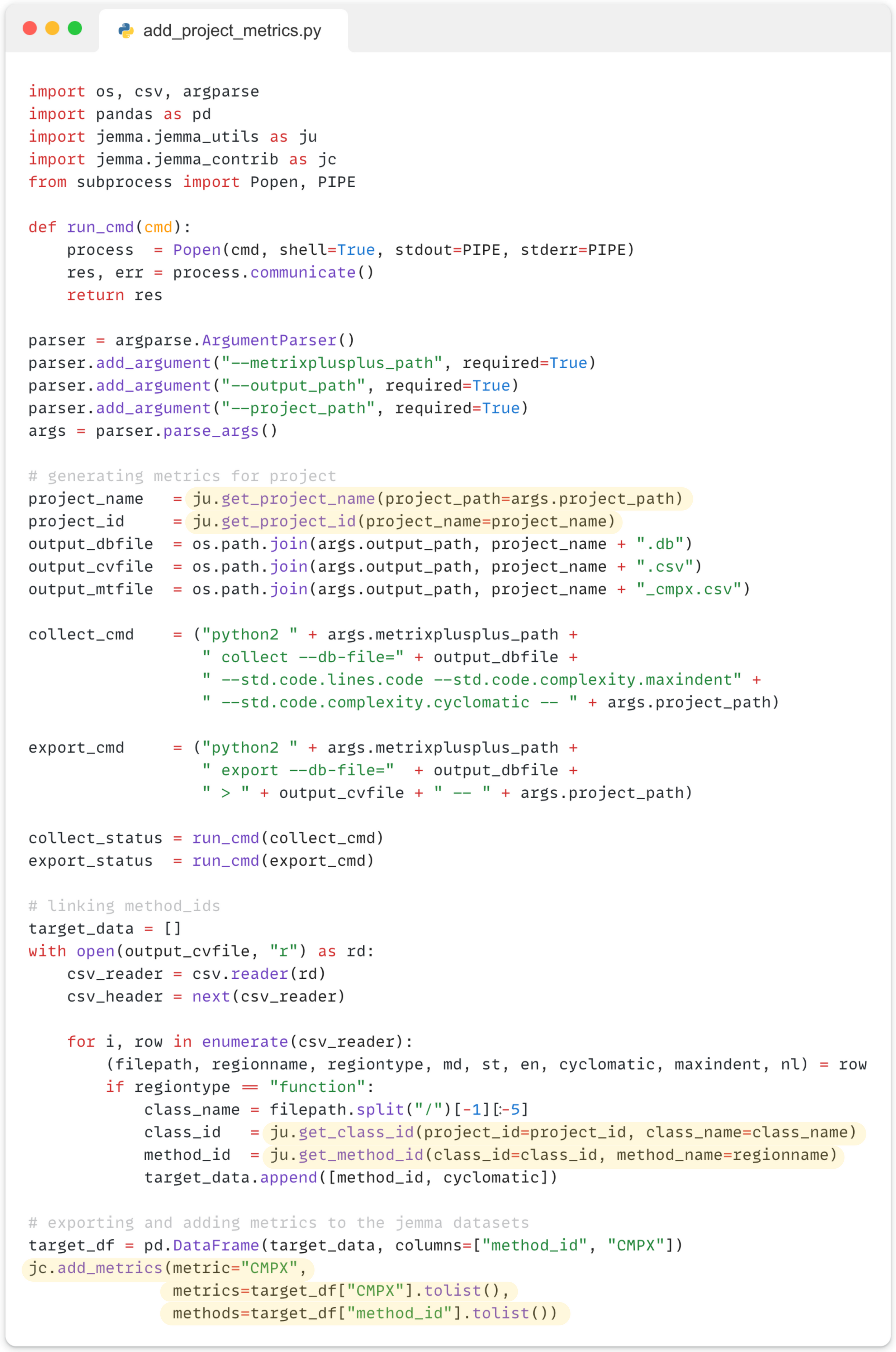}} \\
\textbf{Snippet A.1} Defining and adding new property to \dataset: This snippet shows how to add metrics from the \textit{Metrix++} code analysis tool as a new property to the \dataset dataset.
\label{fig:appendix:define-property}    
\end{figure}

\noindent
The snippet above shows how output metrics from the \textit{Metrix++} tool can be associated with the methods in \dataset and added to the dataset as properties. The yellow highlights mark the \textit{Workbench} \texttt{API} calls in the code snippets.

\newpage
\subsubsection{Adding a new representation}
\label{sec:usages:define-representation}
Different machine learning models of code require different source code representations as input. Some models reason over tokenized source code, while other models reason over more complex structures such as ASTs. Each representation comes with its own set of advantages and drawbacks, while one is extremely feature-rich the other is simple, scalable, and practical. Therefore, the work on representations is still an active area of research---as researchers are continuing to develop new source code representations, or improving the present ones, e.g., by augmenting them with further information.    

\dataset makes is quite simple to do both: create new representations, and modify existing ones. There are three main steps to extend \dataset with a representation: a) accessing a set source code entities, b) generating associated representations, and c) merging the representations to \dataset.

The raw source text, or even representations such as the AST, could be accessed directly to produce new representations for associated code entities. 
And with an array of source code representations readily made available for over 8 million code enities, simplifying or augmenting such representations to create others would save a lot of pre-processing time for the users. In addition, newer representations could also be derived from existing representations based on specific model architectures and needs. 

The feature graph representation which we include in our dataset (see Section~\ref{sec:representations:ftgr}) is built upon the abstract syntax tree (\texttt{AST}) of the code, and is extended with a number of additional edges, depicting various inter-relationships between the \texttt{AST} nodes (e.g., data-flow, control-flow, lexical-usage, call-edges among others). In addition to other necessary information such as line numbers and position numbers of every source code token, supplementary information such as token types, node types, are also provided. Thus, with this representation, the detail of data provided for every code entity is comprehensive. 

\label{par:sec-3.3.6-replacement}
Several derivative representations can be created directly from this one representation by choosing the necessary edge types from the feature graph. For example, for models that require the AST representation as input, choosing just the \textit{Child} edges of the feature graph representation would result in the AST representation. Yet for models that reason over the dataflow information, choosing the \textit{LastRead} and \textit{LastWrite} edges of the feature graph would result in a new dataflow-based representation. {Experimenting with these variants is important, to obtain a better understanding of the trade-offs between the kinds of information available, what they can contribute to model performance, and the difficulty of obtaining the information.}

Beyond deriving descendant representations, adding further edge types to the feature graph is always possible---making it even more feature-rich, and \dataset facilitates such extensions by providing the base representations for several million code entities. In a similar manner, the other representations included with \dataset could also be simplified, modified, augmented to create new representations. 

Once new representations are created, they are associated with corresponding source code entities by means of \texttt{UUIDs}. The representations can then be added to \dataset using the \textit{Workbench} \texttt{APIs}---quite similar to that of adding new properties as demonstrated in Snippet A.1.

\subsubsection{Adding a new project}
\label{sec:usages:adding-project}

{With the evolution of source code over time, and the inclusion of new libraries and modern technologies --- source code datasets (especially the ones built for modeling code) must also keep at pace if they are to remain relevant. We built \texttt{JEMMA} on top of the \texttt{50K-C} Dataset of 50,000 compilable Java projects, however, we want it to be extensible. So, we have provided mechanisms to include additional projects into the fold of \texttt{JEMMA}.}

{Adding a new project to \texttt{JEMMA} involves three main steps: 1) forking the jemma repository, 2) generating the meta-data, representations, properties, call-graphs --- by running the relevant scripts, and 3) making a pull request to added the new-generated data. We provide a simple bash script that helps users generate all the relevant data in one go---generating meta-data and cataloging code entities within the project, generating representations, generating properties, and generating project-level call-graphs. Once the data for the new project is ready, users can then make a pull request to append the data to \texttt{JEMMA} \textit{Datasets}. A detailed tutorial is provided in our documentation. Snippet A.2 lists the command-line procedure to add a new project to \texttt{JEMMA}.}

\begin{figure}[H]
    {\includegraphics[width=110mm]{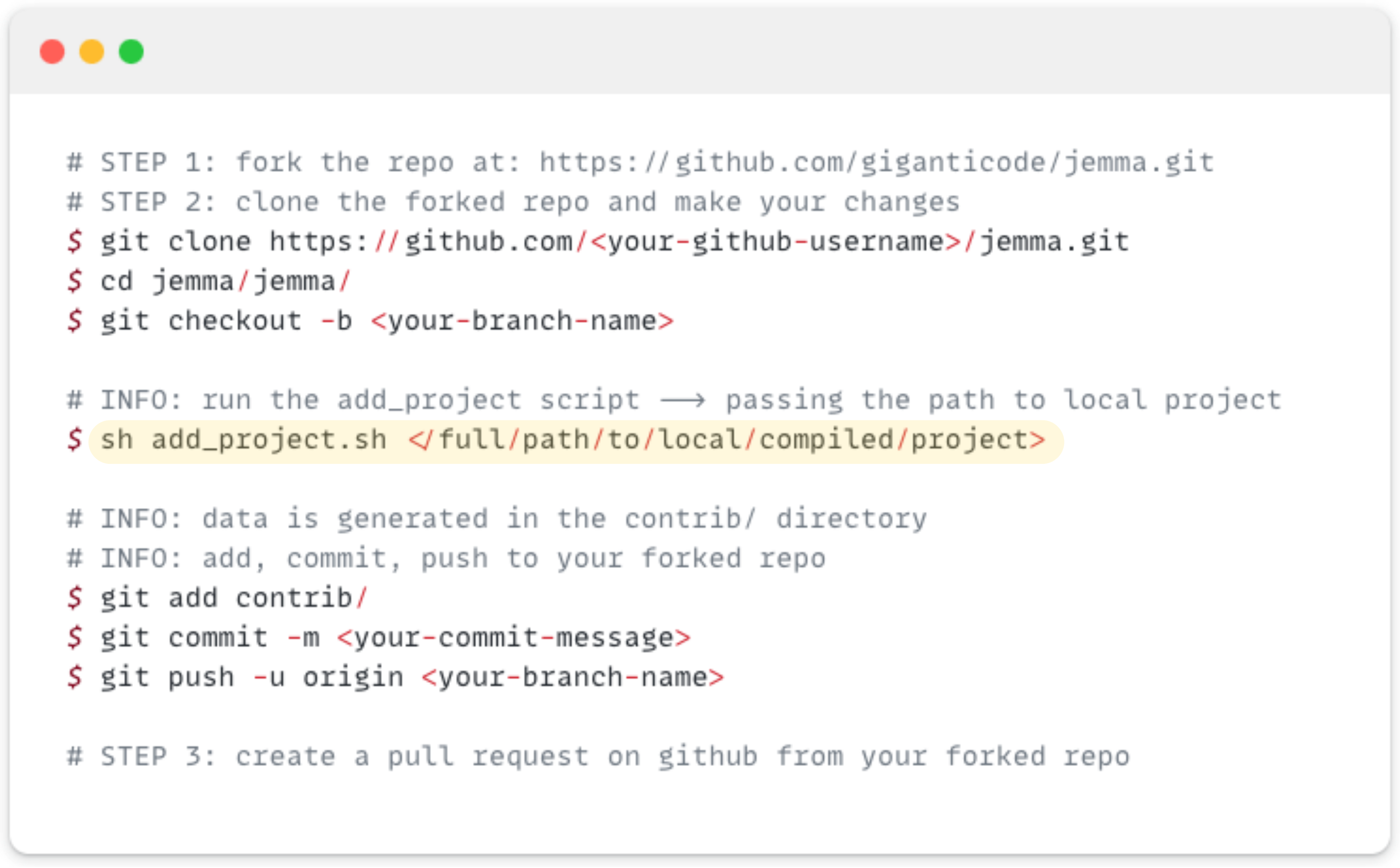}} \\
\textbf{Snippet A.2} Procedure to add a new project to \dataset. \\\texttt{add\_project} runs all the sub-scripts necessary to generate all data for the new project.
\label{fig:add_project}       
\end{figure}

\newpage
\subsection{Using \dataset}
\label{sec:using-jemma}
In this section, we describe various scenarios in which \dataset can be put to use.
In Section~\ref{sec:usages:property-task} we describe how a property can be used to define a prediction task, while discussing ways in which \dataset can help avoid common pitfalls and biases. In Section~\ref{sec:usages:code-task} we explain how source code representations can be used for tasks such as mutation detection and masked prediction. 

In Section~\ref{sec:usages:run-models} we describe how models can be trained and evaluated on prediction tasks using the \textit{Workbench} tools, and finally, in Section~\ref{sec:usages:global-context} we describe how new and extended representations can be formulated with a greater context.

\subsubsection{Defining tasks based on properties}
\label{sec:usages:property-task}

Once a property is defined in \dataset, it can be used in a variety of ways. One such way is to use them as prediction labels for a prediction task. {A good example of such a prediction task is complexity prediction, i.e., given a snippet of code as input, a source code  model must predict its cyclomatic complexity (property) as output.} While this may appear trivial (taking a random sample of entities that have that property defined, and splitting it into training, validation, and test sets), in practice  it is often more complex. This is because care must be taken that the data does not contain biases that provide an inaccurate estimate of model performance. In this context, there are several groups of issues that \dataset helps contend with while defining the task datasets.

\paragraph{Rare data.} The first is that some property values may be very rare, making them hard to learn at all. Examples of this would be uncommon bugs and errors such as, e.g., resource leaks. Since \dataset is large to start with (over 8 million method entities), the scale of data makes it much more likely that there is enough data to learn in the first place, compared to other alternatives.

\paragraph{Defining task labels.} 
\label{par:defining-task-labels}
Once a property is defined, the {\textit{Workbench}} tools provide flexibility in the use of property values as task prediction labels. For instance, for classification tasks, the {\textit{Workbench}} \texttt{API} endpoints allow users to query and obtain a balanced set of prediction labels, ready for training. 

Furthermore, the \textit{Workbench} tools allow the selection of property values as prediction labels that satisfy some criteria based on other properties in the dataset. A subset of the data can also be selected, if one wants to define a task for which data is more scarce, in order to incentivize sample-efficient models. Similar to a Database Workbench, the \dataset \textit{Workbench} allows several such operations in the context of defining prediction labels for a machine learning task, and in managing and retrieving large amounts of specific information.

{Creating datasets for training machine learning models can often be time-consuming and frustrating. Since we start with a lot of data, it might be necessary to filter it down. With the tools included as a part of our \textit{Workbench}, users can query \texttt{JEMMA} and obtain clean, complete, and balanced datasets.}

\begin{figure}[htp]
  \frame{\includegraphics[width=90mm]{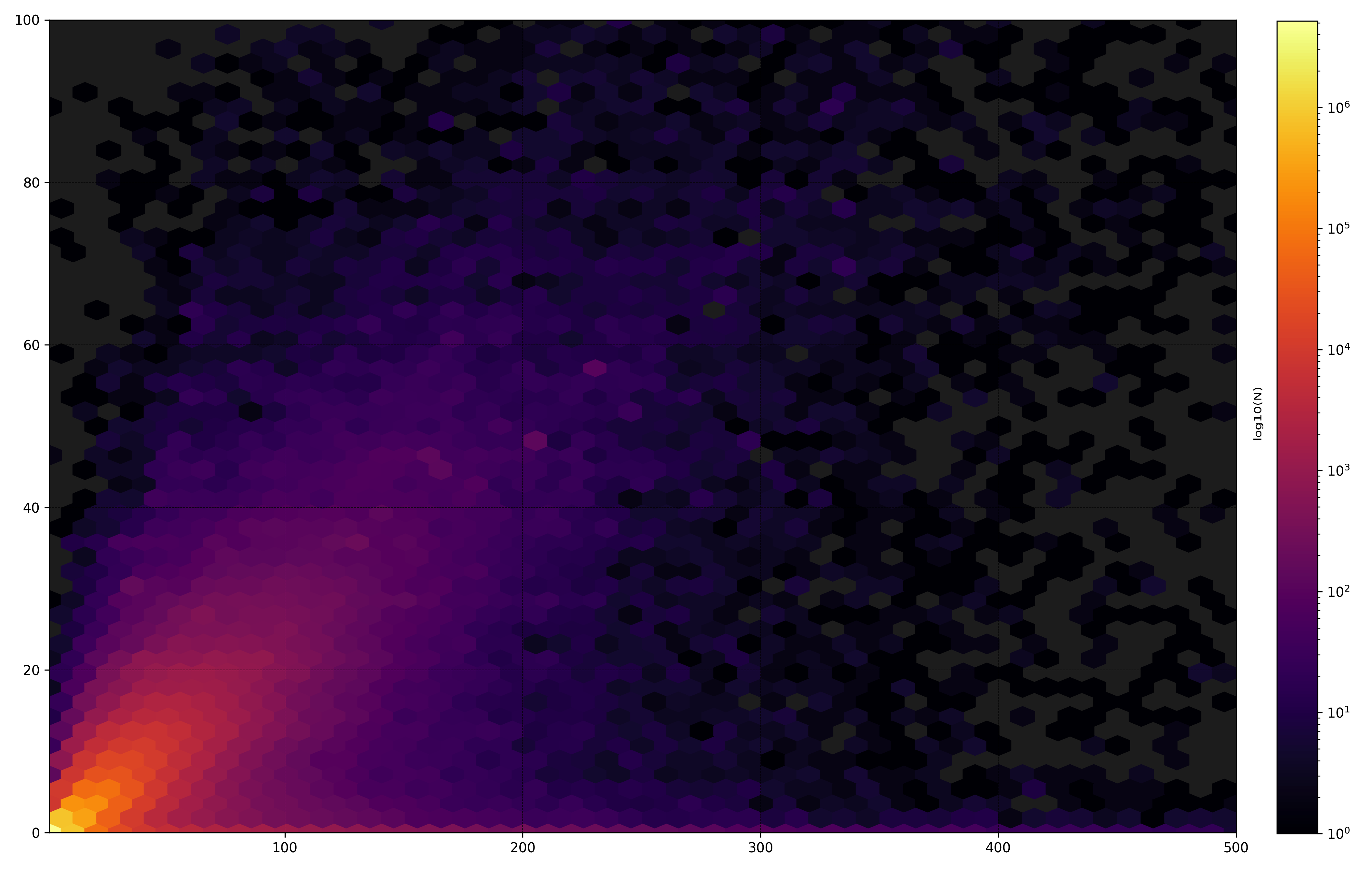}}
\caption{Hexbin plot of cyclomatic complexity (y-axis) vs. source lines of code (x-axis)}
\label{fig:complexity:hexbin}       
\end{figure}

\paragraph{Investigating and mitigating biases.}
\label{par:mitigating-biases}
When defining a task, care must be taken that the models learn from the right signal, and not from the correlated signal that may be easier to learn from, but is not actually a predictive factor. Such issues have been observed in related fields, such as in Computer Vision \cite{beery2018recognition} and NLP \cite{mccoy2019right}, \cite{gururangan2018annotation}. 

In source code, other issues might be present, e.g., a random sample of methods may contain a lot of small methods (including many easy to learn getters and setters), which may inflate performance. For instance, the performance of method naming models is much higher on very short methods (3 lines), than it is for longer methods \cite{alon2018code2seq}. {To mitigate this, \texttt{JEMMA} \textit{Workbench} tools can be used to filter and leverage the already existing properties to empirically investigate the performance of models on the tasks and get insights.} 

In the case of the code complexity example, Figure~\ref{fig:complexity:hexbin} shows the relationship between the size of methods (\texttt{SLOC}) and their complexity as a hexbin plot. We observe that there is an overall tendency for shorter methods to be less complex, and longer methods to be more complex. On the other hand, there also methods that are very long, but have very low complexity (along the bottom axis). This information can be used to properly balance the data, for instance, by making sure that examples that are short and complex, and examples that are long and simple, are also included in the training and evaluation datasets.

\label{par:avoiding-data-leakage}
\paragraph{Avoiding data leakage.} Multiple studies have shown that code duplication is prevalent in large source code corpora \cite{schwarz2012often}, \cite{lopes2017dejavu}, and that it impacts machine learning models \cite{allamanis2019adverse}. {Since \texttt{JEMMA} is built on top of \texttt{50K-C}, we benefit from its selection of projects, which intentionally limited duplication. 50K-C's filtering significantly reduces the risk of leakage across projects.}

{However, since source code can also be repetitive within projects, it could also be a potential source of data leakage. Models that are trained and tested with files from the same project can see their performance affected }\cite{leclair2019recommendations}{. Since \texttt{JEMMA} keeps the metadata of which project a method belongs to, it is easy to define training, validation and test splits that all contain code from different projects, if necessary.}

\subsubsection{Defining tasks based on representations}
\label{sec:usages:code-task}

\dataset can also be used to define tasks that operate on the source code representations themselves, rather than predicting a source code property. These tasks are usually of two forms: a) masked code prediction tasks, and b) mutation detection tasks.

\begin{enumerate}[label=(\alph*)] 
    \item \textbf{Masked code prediction tasks.} In a masking task, one or more parts of the representation are masked with a special token (e.g., "\texttt{<MASK>}"), and the model is tasked with predicting the masked parts of the representations. Examples of this would include the method naming task, where the name of the method is masked, or a method call completion task, where a method call is masked in the method's body. A simpler variant of this would be to use a multiple-choice format, where the model has to recognize which of several possibilities is the token that was masked. \\
    
    \item \textbf{Mutation detection tasks.} In a mutation detection task, the representation is altered with a fixed probability, presumably in a way that would cause a bug (for instance, two arguments in a method call can be swapped \cite{pradel2018deepbugs}). The task is to detect that the representation has been altered. This can either be formulated as a binary classification task (altered vs not altered), or, as a ``pointing'' task, where the model should learn to highlight which specific portion of the given input was altered \cite{vasic2019neural}.
\end{enumerate}

For both of these tasks, the input representation needs to be modified in some way. \dataset can help with this. {For simple modifications (e.g., masking the first occurrence of an operator), it is enough to directly change the default textual representation, and then use the \textit{Workbench} \texttt{APIs} to re-generate the other representations.} Snippet A.3 shows an example of how to generate new representations for a masking task---method call completion.

When doing these kinds of changes, particular care has to be given to data leakage issues. For instance, for a method naming task, the name of the method should be masked in the method's body if it occurs there. Other bias issues can affect these tasks as well, such as a method naming task that over-emphasizes performance on getters and setters. \dataset can be used to analyse the performance of the models on the task and extract insights that may affect the design of the task.

\begin{figure}[H]
  {\includegraphics[width=120mm]{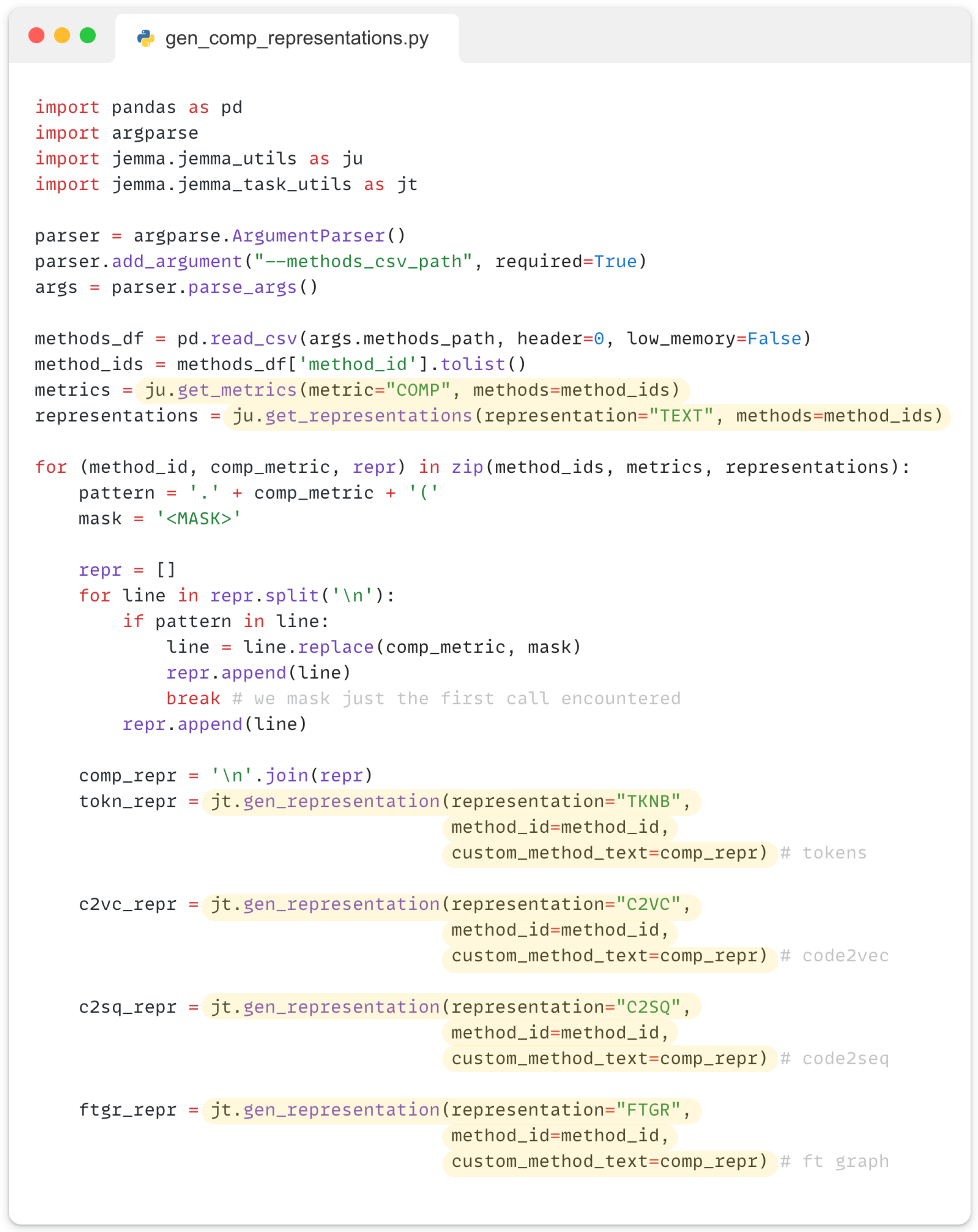}} \\
\textbf{Snippet A.3} Generating new representations for a masking task: This example shows how to generate new representations for a masking task.
\label{fig:appendix:completion}  
\end{figure}

The snippet above shows how representations can be re-generated for masked code snippets. The \texttt{gen\_representation} call handles running all the necessary tools in the background, such that given any source code snippet, representations can be generated on the fly.

\newpage
\subsubsection{Running models}
\label{sec:usages:run-models}
Once a task is defined, the \dataset \textit{Workbench} \texttt{APIs} make it easy to run supported models on the task. Several basic baselines are pre-implemented, and models hosted on the huggingface\footnote{\url{https://huggingface.co/models}} platform are supported out of the box.

\begin{figure}[H]
  {\includegraphics[width=120mm]{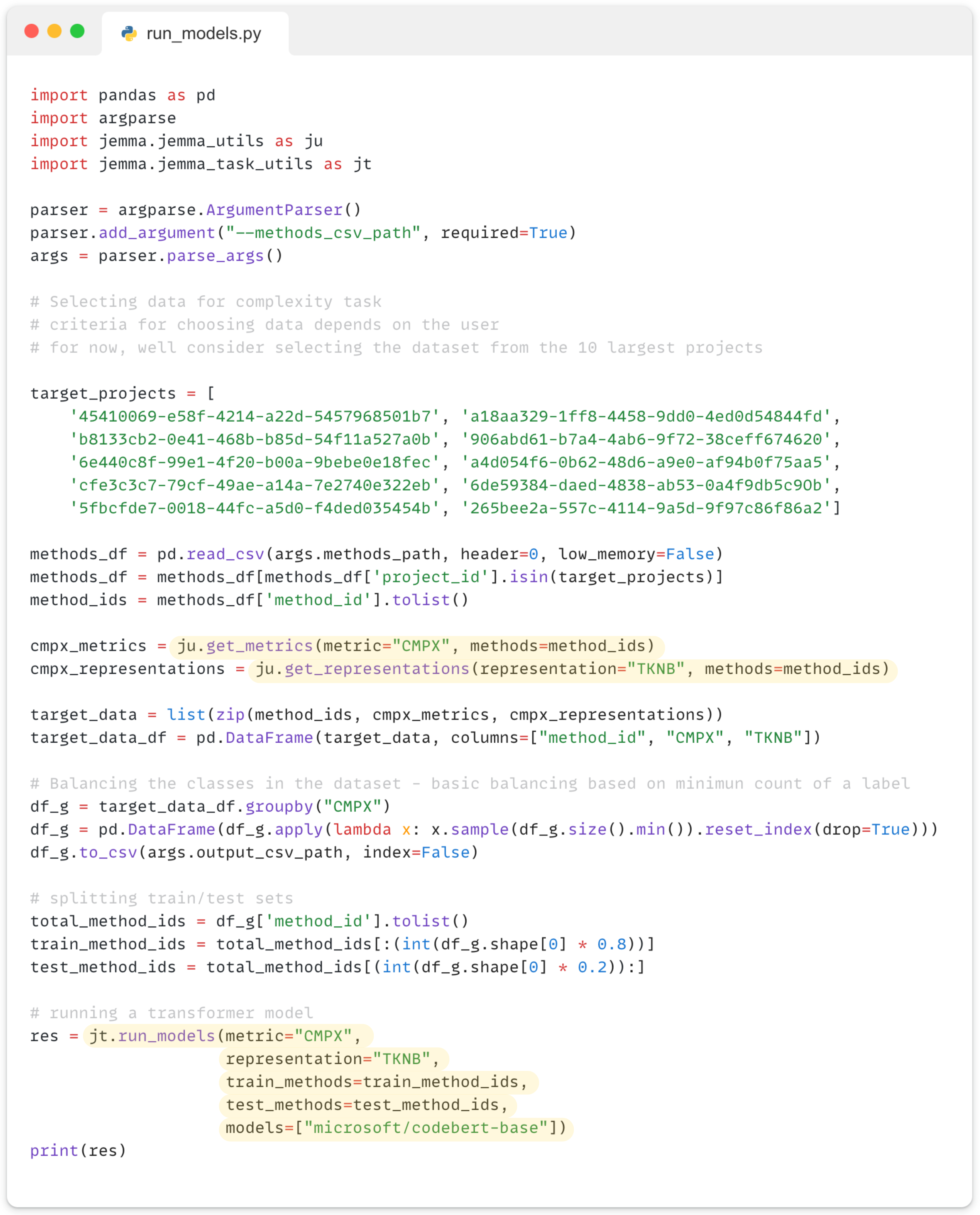}} \\
\textbf{Snippet A.4} Running a Transformer model: Evaluating a transformer model on a prediction task, specifically the cyclomatic complexity prediction task. 
\label{fig:appendix:transformer}       
\end{figure}

\newpage
The \dataset \textit{Workbench} \texttt{API} also facilitates the interaction with other libraries, in particular to run models using the code2vec and code2seq architectures, as well as Graph Neural Network \texttt{(GNN)} models implemented with the ptgnn\footnote{https://github.com/microsoft/ptgnn} library.

Finally, since models based on the transformer architecture, currently, have been the state of the art for a variety of tasks, \dataset allows to easily interface with HuggingFace's Transformer library \cite{wolf2019huggingface}. This allows a variety of pre-trained models to be fine-tuned on the tasks defined with \dataset (such as \texttt{CodeBERT} \cite{feng2020codebert}, \texttt{GraphCodeBERT} \cite{guo2020graphcodebert} etc.). Snippet A.4 shows how to run a Transformer model on the method complexity task using the \textit{Workbench}.

\subsubsection{Defining representations with larger contexts}
\label{sec:usages:global-context}
One of our goals with \dataset is to allow experimentation with novel source code representations. In particular, we want users to be able to define representations that can take into account a larger context than a single method, or a single file, as is done with the vast majority of work today.

The key to building such extended representations is to have access to the necessary contextual information. The extensive pre-processing we did to create \dataset gives us all the relevant tools to gather that information. The metadata of \dataset documents the containment hierarchies (e.g., which files belong to which project, and which classes belong to which package etc.) and provides the ability to uniquely and unambiguously identify source code entities at different granularities. In addition, the call graph data documents which are the immediate callers and callees of each individual method. Since the call graphs link to each method identified by their \texttt{UUID}, all the properties of the methods, including their representations, can be accessed easily and systematically. Thus, from navigating the call graph and the containment hierarchy, various types of global contexts can be defined at the class-, package-, or even project-level. We present two simple examples in the appendix. 

Snippet A.5 (a) and Snippet A.5 (b) show how to combine representations of a given method with the representations of its direct callees to include greater context. We encourage users to experiment with more complex representations adding context information that go beyond a single method. The extensive pre-processing of data, at the scale of tens of thousands of projects, combined with the \textit{Workbench} makes it possible to do so easily.

\begin{figure}[H]
  \captionsetup{justification=centering}
  {\includegraphics[width=110mm]{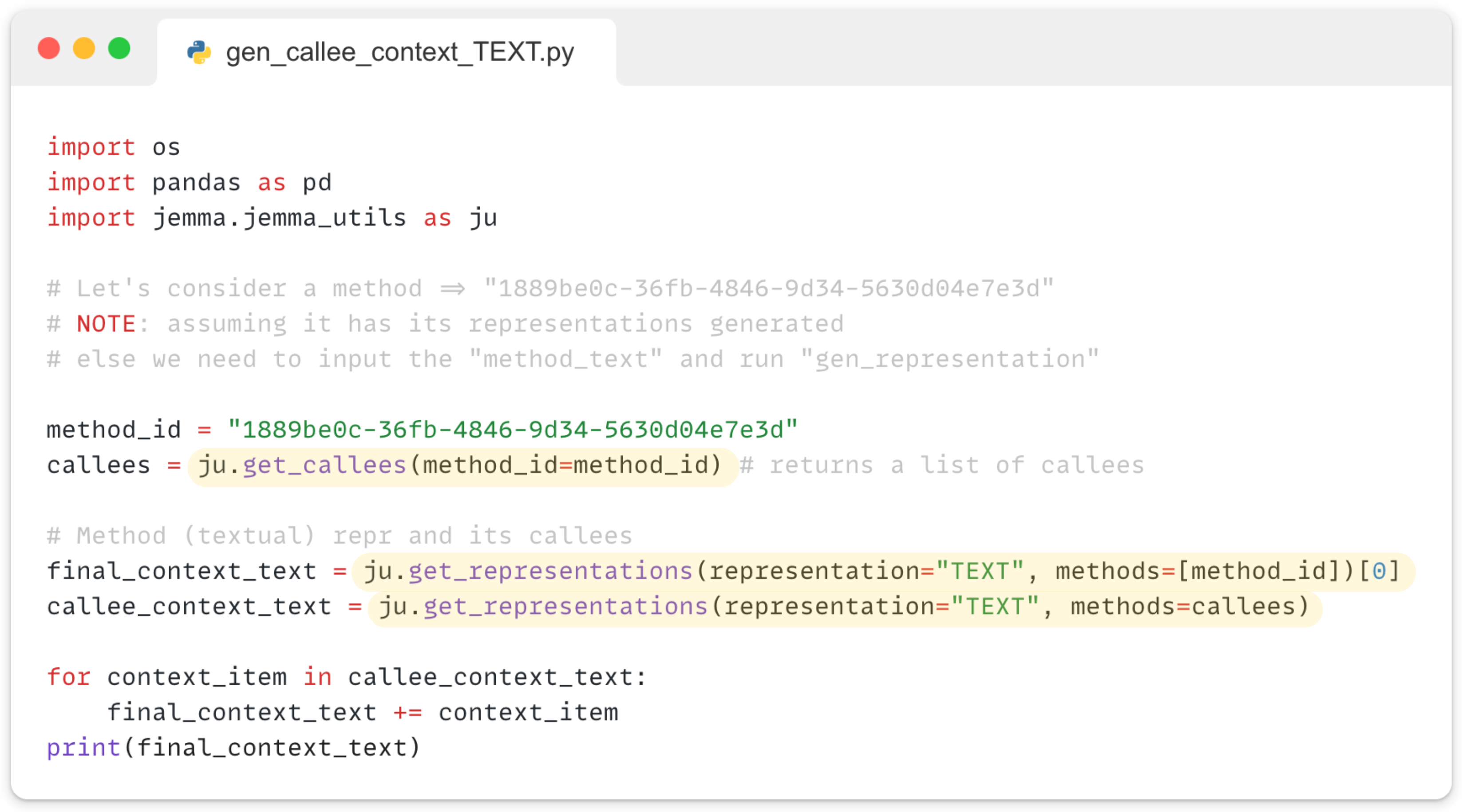}} \\
\textbf{Snippet A.5 (a)} Building a Context: This example shows how to combine a textual representation of a method with additional context from its direct callees.
\label{fig:appendix:text-context}  
\end{figure}

\begin{figure}[H]
  {\includegraphics[width=110mm]{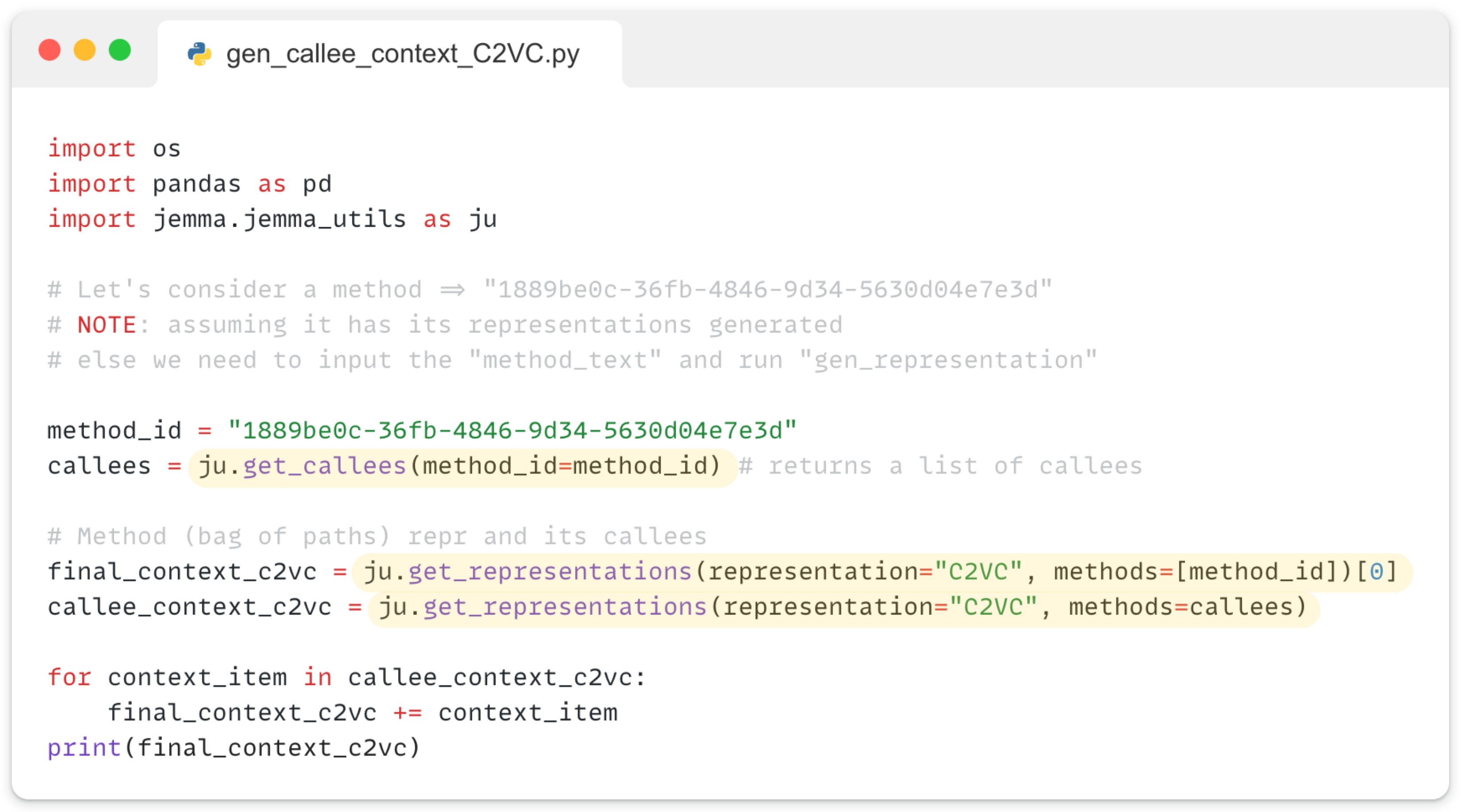}} \\
\textbf{Snippet A.5 (b)} Building a Context: This example shows how to combine a 
code2vec representation of a method with additional context from its direct callees.
\label{fig:appendix:path-context}    
\end{figure}

\section{{Empirical Study I: On the extent of non-localness of software}}
\label{sec:non-localness}
Since software is made of a lot of interacting entities across files, packages, and projects---modeling source code by learning on smaller entities of code (e.g., methods) can at best provide a localized understanding of source code for a given task. We hypothesize that in order to improve our source code models comprehensively beyond localized understanding, we must include non-local contextual features while modeling source code. As a result, we must study the extent of non-localness of software and whether non-local context is useful. 

\noindent
{In this section, we demonstrate the utility of \texttt{JEMMA} \textit{Dataset} and \textit{Workbench} through an empirical investigation.} We study the extent to which software is made up of interacting functions and methods in a sample of projects contained in \dataset by analysing their call graphs. We observe how often method calls are local to a file, cross file boundaries, or are calls to external {APIs}. Then, we analyze the performance on the method call code completion task through the lens of call types when non-local context is added. We pose the following research questions for this part of our study. 

\begin{itemize}
    \item[$\circ$] {RQ. 1. To what extent are method calls non-local?}
    
    \item[$\circ$] {RQ. 2. What is the effect of adding non-local context for the method call completion task?}
\end{itemize}

\subsection{Extent of non-localness of code}
\label{sec:non-localness:extent}

{Since methods generally do not exist in isolation, a large number of associations can be found among source code entities in project-wide contexts. Thus, we pose the first research question to determine the extent of non-local association of methods with other source code entities defined the project and beyond. In other words, we attempt to determine the extent of non-localness dependence of software.} 

{To determine the extent of non-localness of software, we first track the interacting method entities within projects. For each method defined in a project, we count the number of unique callers and number of unique callees in the call graph and find that over 70\% of the methods have at least one unique caller, and over 72\% of the methods have at least one unique callee. This confirms that software is highly interconnected.}

{But to what extent is the interconnected-ness strictly non-local?} To study this
we measure the frequency of the various types of method calls in these projects. We classify all calls into four categories as listed below:

\begin{itemize}
    \item[$\circ$] \emph{Local calls.} The entity is defined in the same file; thus, a machine learning model that has a file context would be likely to see it.
    \item[$\circ$] \emph{Package calls.} The entity is defined in the same Java package (i.e., the classes as in the same file directory).
    \item[$\circ$] \emph{Project calls.} The called entity is defined in the project, but in a different package than the caller.
    \item[$\circ$] \emph{API calls.} The called entity is not defined in the project, but is a call to an imported library.
\end{itemize}

\begin{figure}
  {\includegraphics[width=90mm]{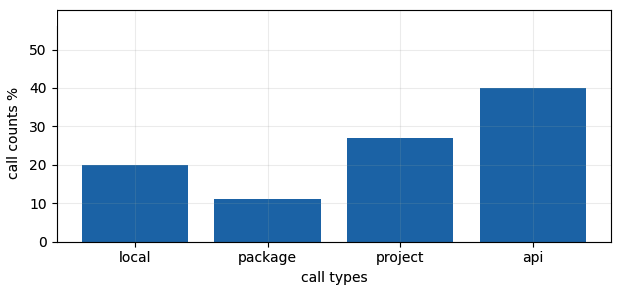}}
\caption{Distribution of calls by type}
\label{fig:calls-by-localness}       
\end{figure}

Figure~\ref{fig:calls-by-localness} shows the distribution of the calls. We can see that only 20\% of calls are \emph{local calls}; these are the calls whose callees are visible to the models that learn from the entire file context, such as CoCoGum \cite{wang2020cocogum}; the remaining 80\% of calls are \emph{non-local} and are not visible to models that learn from the file context only. Of these, 12\% are \emph{package calls}; thus a model that builds a context of the classes in the same directory to absorb a larger context than the file would have visibility into these callees. On the other hand, 28\% of calls are \emph{project calls}, thus models would need either a larger context, or the ability to select from this larger context in order to have visibility in the callees. %
Finally, \emph{API calls} constitute 40\% of all calls (inflated by the vast majority of standard library calls). While these are out of reach for most models, a silver lining is that, in practice, it is often possible to learn common \textit{API} call usages as modern large-scale source code models do.

\subsection{Impact of non-localness on code completion}
\label{sec:non-localness:completion}
Having some insight into the extent of non-localness of software, we now look at whether adding non-local context can have an impact on model performance.

The study of \citet{hellendoorn2019code} investigated the differences between code completions from synthetic benchmarks and real-world completions. Although all forms of benchmarks are useful \cite{karmakar2019establishing}, it found that synthetic benchmarks underweighted the frequencies of method completions relative to real-world completions, and that those were the most difficult. They also observed that among method completions, the hardest ones were the ones from project internal identifiers. Hellendoorn's study offers valuable insights but has limitations: the data for the real-world completions was relatively small (15,000 completions); the study evaluated \texttt{RNNs} with a closed vocabulary, which were unable to learn new identifiers. Since then, open-vocabulary models \cite{karampatsis2020big} \cite{ciniselli2021empirical} have considerably improved the state of the art.

\label{para:hellendoorn-motivation}
{The key point from }\citet{hellendoorn2019code}{ that motivates our study is the observation: \textit{code completion models struggle the most with project internal identifiers, e.g. method-call completions}. This is because models such as \texttt{RNNs} have a very limited context size, so they are unable to know which identifiers are defined in the project. And since this information is spread over the entire project, it motivates our choice to design a code completion task with much more data-points that focuses particularly on method-call completions considering a project-wide context.}

\noindent
We use the \dataset \textit{Workbench} to analyse the performance of three state-of-the-art Transformer code models, with the natural-language \texttt{BERT} model as the baseline, on a derivative of the code completion task: method-call completion.

We first train our models {without} any additional context; and report the exact-match\footnote{We intentionally chose against string similarity-based metrics, since the completions need to be \emph{exactly} the same as the \textit{APIs} or method-names actually defined in the project.}  accuracies for method-call predictions. We then train our models including context from caller-callee method entities defined across the project, comparing the results. A brief explanation on how larger contexts may be constructed is presented in the Appendix~\ref{appendix:a1}. For our experiment, we have considered the context information from a method's 1-hop neighborhood, considering all possible callee names.  Furthermore, informed by the observations from the previous section, we separately analyse the performance of models on different strata of the test set, according to the categories defined above: \emph{local calls}, \emph{package calls}, \emph{project calls} and \emph{API calls}. 

\begin{table}[t]
\label{tab:completion:types}
\caption{{Method call completion (by call types) without vs. with context, \% improvement.}  \\ 
        Scores for: \texttt{A - {BERT}, B - {CodeBERTa}, C - {CodeBERT}, D - {GraphCodeBERT}} } 

    \resizebox{\textwidth}{!}{%
    \begin{tabular}{|l|
    c|>{\columncolor[gray]{0.9}}c|c|
    c|>{\columncolor[gray]{0.9}}c|c|
    c|>{\columncolor[gray]{0.9}}c|c|
    c|>{\columncolor[gray]{0.9}}c|c|}
        \hline
        \multicolumn{1}{|c|}{\multirow{2}{*}{}}                                        
            & \multicolumn{3}{c|}{\textit{Local}}                         
            & \multicolumn{3}{c|}{\textit{Package}} 
            & \multicolumn{3}{c|}{\textit{Project}}
            & \multicolumn{3}{c|}{\textit{API}} \\ 
         
        \hline %
        \multicolumn{1}{|c|}{}                 
            & \texttt{n/c} & \texttt{c}  & $\pm$
            & \texttt{n/c} & \texttt{c}  & $\pm$
            & \texttt{n/c} & \texttt{c}  & $\pm$
            & \texttt{n/c} & \texttt{c}  & $\pm$
            \\ \hline

\multicolumn{1}{|l|}{\texttt{A}}
 &  0.102  &  0.159  &  \textcolor{teal}{ 56\%}    &  0.112  &  0.154  &  \textcolor{teal}{ 37\%}    &  0.144  &  0.181  &  \textcolor{teal}{ 26\%}    &  0.296  &  0.336  &  \textcolor{teal}{ 14\%}    \\ \hline
\multicolumn{1}{|l|}{\texttt{B}}
 &  0.171  &  0.284  &  \textcolor{teal}{ 66\%}    &  0.255  &  0.362  &  \textcolor{teal}{ 42\%}    &  0.278  &  0.370  &  \textcolor{teal}{ 33\%}    &  0.524  &  0.606  &  \textcolor{teal}{ 16\%}    \\ \hline
\multicolumn{1}{|l|}{\texttt{C}}
 &  0.137  &  0.187  &  \textcolor{teal}{ 36\%}    &  0.144  &  0.176  &  \textcolor{teal}{ 22\%}    &  0.182  &  0.209  &  \textcolor{teal}{ 15\%}    &  0.376  &  0.397  &  \textcolor{teal}{ 6\%}    \\ \hline
\multicolumn{1}{|l|}{\texttt{D}}
 &  0.142  &  0.188  &  \textcolor{teal}{ 32\%}    &  0.147  &  0.176  &  \textcolor{teal}{ 20\%}    &  0.184  &  0.209  &  \textcolor{teal}{ 14\%}    &  0.379  &  0.398  &  \textcolor{teal}{ 5\%}    \\ \hline

    \end{tabular}}
\end{table}

\paragraph{Task definition:}
\label{par:task-definition}
We define the method call completion task as a masking task: for each method snippet, we mask one single method call in the code snippet at random. These methods can be present in the same class (18\% of the dataset), in another class in the same package (10\%), in another package in the system (26\%), or imported from a dependency (46\%). The goal of the task is for a source code model to predict the \emph{exact} method name that was masked. {We sample 100K methods from the \texttt{JEMMA} \textit{Datasets}, splitting 80K samples as training data, 5K as validation data, and 15K as test data for training and evaluation.}

\paragraph{Model Performance:}
\label{par:model-performancex}
We analyze the performance of three large-scale Transformer models of code: \texttt{CodeBERTa}, \texttt{CodeBERT}, and \texttt{GraphCodeBERT}. We use th \texttt{BERT} model as the baseline model for this task. {All of these models accept sequences of tokens as input, so we use the token representation for training.}

{Table 3 shows the accuracy across the call types when the models were trained without context and with additional context. We observe an improvement across all models, and across all call-types, when additional context was included. This shows that tasks like method-call completion---an integral component for the success of code completion, rely on the information beyond the local context and could benefit from additional project-wide context.} 

\noindent
Furthermore, we can clearly see that accuracies are much higher for the \emph{API calls} than the other categories, with the second highest being the \emph{local calls}; the \emph{project calls} and the \emph{package calls} having the lowest performance. While we can expect that different models would perform differently, the margin between \textit{API calls} and the other types of calls is wide enough to demonstrate that the models perform much better at predicting \textit{API calls} than calls defined within the project.  \vspace{-12pt}

\subsection{Implications}
\label{sec:non-localness:implications}
From the observations in the previous sections we see that: a) {a large number of method calls are non-local, i.e., collectively 80\% of the method calls are {not} local to the same parent class,} 
and b) source code models struggle to predict call completions of methods defined in the same project, but improve when additional non-local context is added.

This nudges us to explore the notion of designing and training source code models in a way that it can reason over a larger context of information, at least at the project-level. It becomes necessary to determine ways in which models could be made aware of the inter-relationships that exist among code entities by providing a feature-rich representation with as much context information that we can possibly fit. With the depth and extent of data that we have gathered, and with the help of our \textit{Workbench}, users can easily construct extended contexts beyond the method-level for use in training context-aware source code models in the future.

\section{{Empirical Study II: OOW is the next OOV}}
\label{sec:OOW-OOV}
The studies in the previous section show that software entities have inter-relationships which when considered can affect the performance of models. This section provides data to inform the design of possible model architectures that can absorb a larger context. In particular, we focus on the \emph{size} of this context, as deep learning models can be strongly affected by the input size. 

Machine learning models of code once struggled with \emph{Out-Of-Vocabulary} (\texttt{OOV}) issues \cite{hellendoorn2017deep}, until more recent models introduced and adopted an open vocabulary \cite{karampatsis2020big}.

We argue that the next problem to address is the \emph{Out-Of-Window} (\texttt{OOW}) issue: all modern state-of-the-art models tend to have a fixed input size, which may not be enough to fit the additional context needed. How to best use this limited resource is thus, an open problem. To that effect, we pose the following research questions in this section:

\begin{itemize}
    \item[$\circ$] {RQ. 1.  Given the need for fitting additional context, are English-based model tokenizers comparable to language-specific tokenizers?} 

    \item[$\circ$] {RQ. 2.  From the perspective of context size, what types of code entities fit modern transformer models at different input size limits?} 
\end{itemize}

\subsection{Transformers, window sizes, and tokenizers}
\label{subsec:tokenizers}

\newcommand\entok{{\texttt{RoBERTa tokenizer}}\xspace}
\newcommand\codetok{{\texttt{CodeBERTa tokenizer}}\xspace}
\newcommand\javatok{\texttt{Java BPE tokenizer}\xspace}
\newcommand\lexitok{\texttt{Java Parser}\xspace}

\newcommand\trsmall{\emph{Small}\xspace}
\newcommand\trbase{\emph{Base}\xspace}
\newcommand\trlarge{\emph{Large}\xspace}
\newcommand\trxl{\emph{XL}\xspace}
\newcommand\trxxl{\emph{XXL}\xspace}

For many machine learning tasks, Transformer-based models \cite{vaswani2017attention} are now the state of the art. Some transformer models that have achieved state-of-the-art performance on source code tasks include \texttt{CodeBERT} \cite{feng2020codebert}, \texttt{CodeBERTa} \cite{DBLP:journals/corr/abs-1910-03771}, \texttt{PLBART} \cite{ahmad2021unified}, \texttt{CodeT5} \cite{wang2021codet5}, \texttt{CodeGen} \cite{nijkamp2022conversationalcodegen}, \texttt{GraphCodeBERT} \cite{guo2020graphcodebert}. \texttt{Codex} \cite{chen2021evaluating} is yet another of these large pre-trained Transformer models, that has demonstrated compelling competence on a variety of tasks without necessarily needing fine-tuning, ranging from program synthesis, program summarization \cite{chen2021evaluating}, to even program repair \cite{prenner2021automatic}. 

However, all Transformers that follow the classic architecture have fixed window sizes: for \texttt{CodeBERT}, it is 512 tokens, while for the largest Codex model (codex-davinci), it is 4,096 tokens. If an input is longer than the window, it is generally truncated. 
Transformers rely on self-attention, where the attention heads \textit{attend} to each pair of tokens: the complexity is hence quadratic, which renders very large windows prohibitive in terms of training time and inference time. This raises the question: for a given window size, how much code can we expect to fit?

Since Transformers are open-vocabulary models, the tokens that they take as input are actually subtokens, common subsequences of characters learned from a corpus, rather than entire tokens. A word that would be unrecognized by a closed-vocabulary model will, instead, be split up in several more common subtokens. This means that the number of lexical tokens in a method does not match the length of the method in terms of subtokens, and depends on the corpus that was used to train the subword tokenizer. It is important to note that both \texttt{CodeBERT} and \texttt{Codex} are not models trained from scratch on source code: given the amount of time needed to train such a model from scratch, previous models trained on English (\texttt{RoBERTa} for \texttt{CodeBERT}, a version of \texttt{GPT-3} for \texttt{Codex}) were fine-tuned on source code instead. This means that both \texttt{CodeBERT} and \texttt{Codex} use a subword tokenizer that was not learned for source code, but for English, which might lead to sub-optimal tokenization.

To estimate the number of tokens that a method will take in the model's input window, we first selected a sample of 200,000 Java methods from \dataset, and used several subword tokenizers to estimate the ratio of subtokens that each subword tokenizer will produce. We first noticed that the choice of subword tokenizer has a significant impact on the produced tokenization, and consequently the amount of code that can fit in a model's input window. We used the following tokenizers for our analyses:

\begin{itemize}
\setlength\itemsep{0.2em}
    \item[$\circ$] \entok. A byte-level BPE tokenizer, trained on a large English corpus, with a vocabulary of slightly more than 50,000 tokens. A similar tokenizer is used by \texttt{CodeBERT} and \texttt{Codex}.

    \item[$\circ$] \codetok. The tokenizer used by \texttt{CodeBERTa}. This tokenizer was trained on source code from the CodeSearchNet corpus, which comprises of 2 million methods in 6 programming languages, including Java. 
    
    \item[$\circ$] \javatok. A tokenizer similar to \codetok, trained on 200,000 Java methods from Maven, instead of several languages. 

    \item[$\circ$] \lexitok. A standard tokenizer from a Java Parser, that does not perform sub-tokenizations. We use this as a baseline for our analyses. 
\end{itemize}

\paragraph{}
\label{par:calculate-85-percent-abc}
\vskip 1mm%
{We tokenized Java source code using the tokenizers above, keeping the \texttt{Java Parser} (standard tokenizer) as the baseline, and then calculated the average percentage- increase or decrease in the number of generated tokens.} {The \texttt{CodeBERTa tokenizer} learned on multiple programming languages, on average, generates 98 tokens per 100 tokens of the baseline \texttt{Java Parser} tokenizer.} This is expected since some common token sequences can be merged in a single token (e.g, \textit{();} can be counted as one token instead of three tokens). The learned \javatok is even more efficient, using on average 85\% of the tokens {(i.e. it generates 85 tokens per 100 tokens of the standard tokenizer)}. This is possible since, for instance, specific class names will be common enough that they can be represented by a single token (e.g., \textit{ArrayIndexOutOfBoundsException}). On the other hand, the \entok is considerably less efficient, needing 126\% of the lexical tokens compared to the baseline.  

With an equal vocabulary size, the most efficient language-specific encoding can fit close to 25\% more effective tokens in the same window size. For a window size of 512, a Java-specific tokenizer will, on average, be able to effectively fit 602 actual tokens, while the English-specific tokenizer---used by both \texttt{CodeBERT} and \texttt{Codex}---will be able to fit only 409 actual tokens. For example, for tokens such as \textit{ArrayIndexOutOfBoundsException}, efficient language-specific code tokenizers will tokenize it as a single token,
rather than six separate tokens. \vspace{-8pt}

\paragraph{}
\label{par:calculate_85_percent_explain}
{This establishes that language-specific code tokenizers are more efficient in tokenizing source code compared to their English-language counterparts. And since almost all model architectures have a maximum input size limit, the tasks that specifically rely on additional context information can benefit from efficient tokenizers, whereby input source code snippets can be represented in less number of tokens leaving space for additional context information. From this perspective, efficient tokenizers can be helpful because having the possibility of including additional context can ultimately improve model performance.}

\subsection{Fitting code entities}
\label{subsec:6:fitting-code-entities}
Taking the same 400 projects as in the code completion study in Section~\ref{sec:non-localness}, we tokenize the methods and the classes in these projects with the four tokenizers above. We then estimate the size of higher-level entities (packages and projects) by summing the token sizes of the classes in them. We compare these sizes against a range of Transformer window size thresholds:

\begin{itemize}
\setlength\itemsep{0.1em}
    \item[$\diamond$] \trsmall. A window size of 256 tokens, representing a small transformer model 
    \item[$\diamond$] \trbase. A window size of 512 tokens, representing a model with the same size as \texttt{CodeBERT} \cite{feng2020codebert}. 
    \item[$\diamond$] \trlarge. A window size of 1,024 tokens, which is the context size used by the largest \texttt{GPT-2} model \cite{radford2019language}. 
    \item[$\diamond$] \trxl. A window size of 2,048 tokens, which is the context size used by the largest \texttt{GPT-3} model \cite{brown2020language}. 
    \item[$\diamond$] \trxxl. A window size of 4,096, which is the context size used by the largest \texttt{Codex} model \cite{chen2021evaluating}.
\end{itemize}
It is important to note that these models are very expensive to train. In practice, training a model with a \trbase window size of 512 tokens, from scratch, is a significant endeavour inaccessible for most academic groups, leaving fine-tuning as the only practical option. Only industry research groups or large consortiums of academics may have the resources necessary to train such large models. Even conducting inference on the largest of models becomes impractical due to their size.

\begin{figure}[htp]
  \includegraphics[width=120mm]{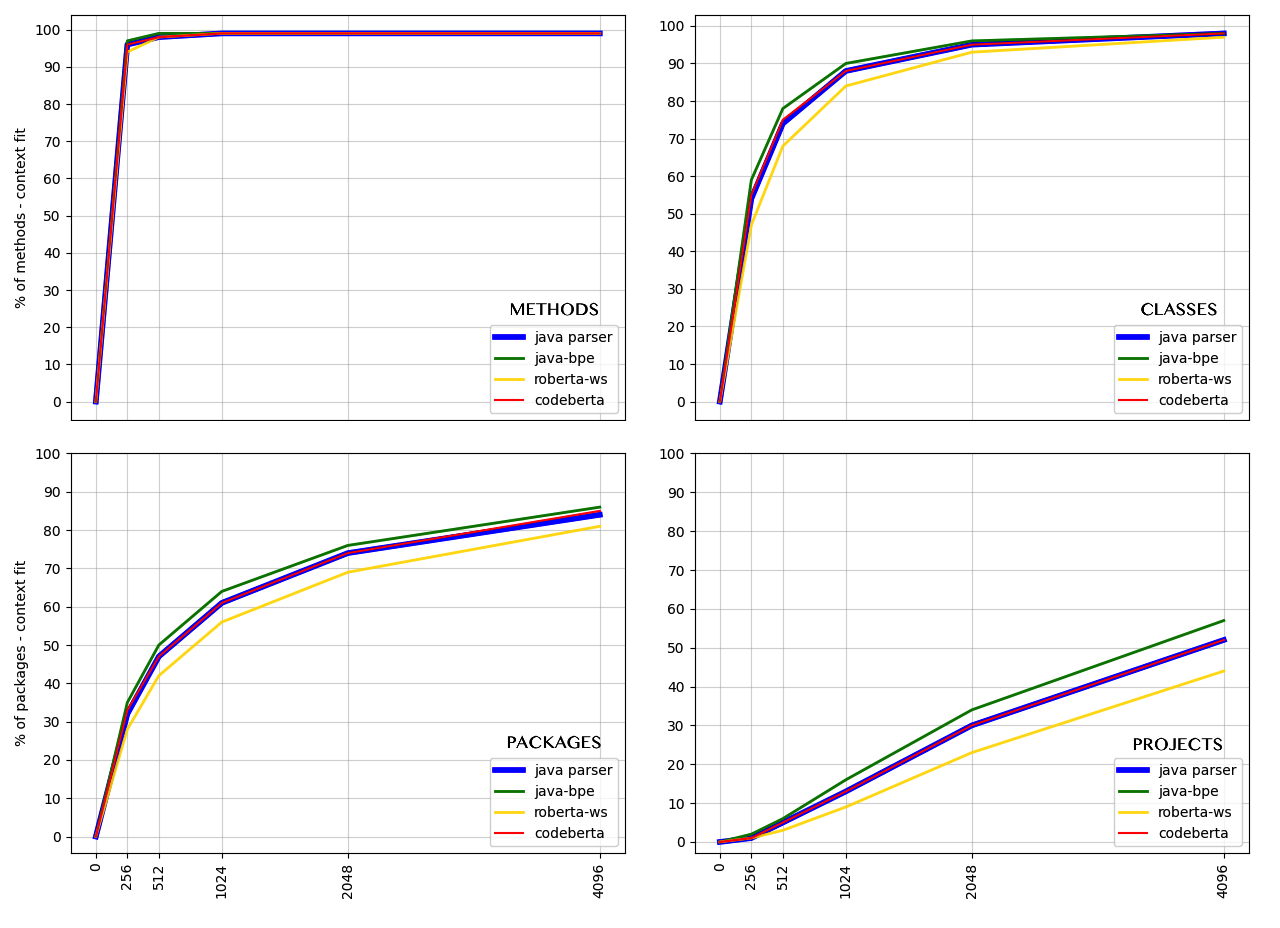}
\caption{Percentage context-fit for: (a) methods; (b) classes; (c) packages; (d) projects.}
\label{fig:all-distribution}       
\end{figure}

\subsubsection{Methods} 
Figure~\ref{fig:all-distribution} (top-left) shows the percentage of methods that fit within different window size thresholds. We can see that even the \trsmall model (with a maximum input size of 256 tokens) is able to comfortably fit the vast majority of methods (over 94\%). The choice of tokenization still matters, as more efficient tokenization can make up to 97\% of methods fit in the \trsmall model. Overall, a \trbase model with a window size of 512 tokens can fit 99\% of the methods in our sample, while only extreme outliers do not fit even in the \trxxl models with a limit of 4096 tokens.

\subsubsection{Classes}
We tokenize the entire source file to compute the context size needed for classes.
Figure~\ref{fig:all-distribution} (top-right) shows the percentage of classes that fit within different window size thresholds. We can see that models with smaller window sizes are beginning to struggle. A \trsmall model with a token limit of only 256 tokens will be able to process between 47-59\% of the classes. A \trbase model would instead be able to process between 68 and 78\% of the classes, while a \trlarge model would fit up to 90\% of the classes. \trxl models can fit almost more than 95\% of the classes on average, but some outliers (2-3\%) will remain even for a Codex-sized model.

\subsubsection{Packages}

Figure~\ref{fig:all-distribution} (bottom-left) shows the percentage of packages that fit within different window size thresholds. Models with smaller window sizes struggle significantly, with a \trsmall model able to fit only a 30 to 35\% of the packages, and a \trbase model 42 to 50\%, depending on the tokenization. A \trlarge model succeeds in 55 to 65\% of the cases. We can clearly see that even the models with the largest token limits start to struggle while fitting packages into context: 69-76\% fit in a window size of 2048 tokens, and 81-86\% fit in a window-size of 4096 tokens.

\subsubsection{Projects}
On average, only half the projects can fit in the window sizes, as seen in Figure~\ref{fig:all-distribution} (bottom-right). But since we expect that larger projects would behave differently, we present a context-fit graph for projects based on size (Figure~\ref{fig:project-distribution-buckets}). 
We observe that while models with large window sizes are able to fit 66-81\% of small-sized projects that have 20 or fewer classes, the rate drops drastically as project size increases---falling to 14-28\% for medium-sized projects. Beyond this, very few (less than 6\%) of the larger projects can fit any window-size.
Of note, the largest projects that do not fit the model window sizes, being the most complex, are likely the ones for which the source code models might be the most useful.

\begin{figure}[htp]
  \includegraphics[width=120mm]{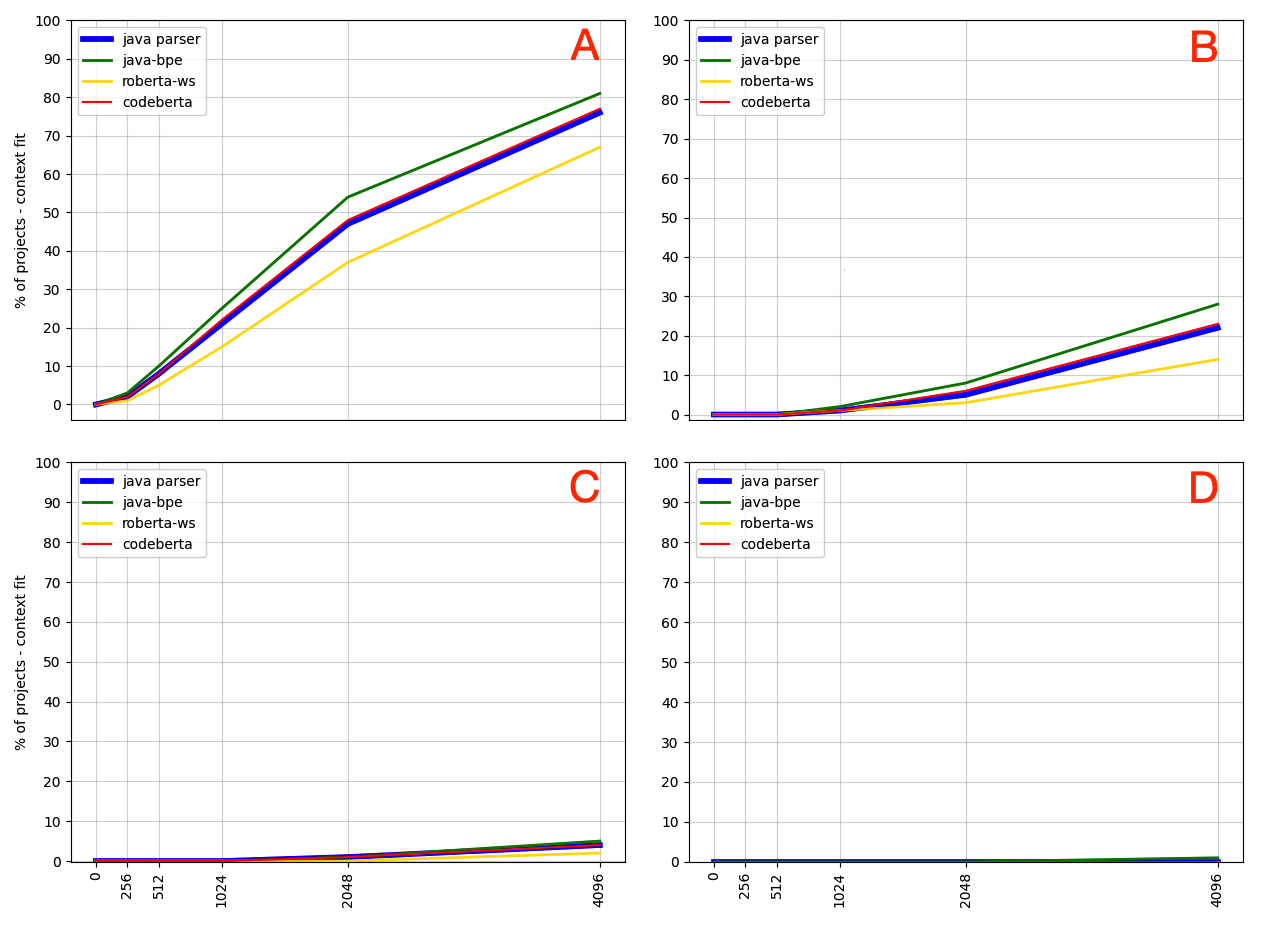}
\caption{Percentage of context-fit for full projects by project sizes. A: up to 20 classes; B: 21-50 classes; C: 51-100 classes, D: more than 100 classes.}
\label{fig:project-distribution-buckets}       
\end{figure}

\subsection{Implications}
\label{subsec:6:implications}
In addressing our research questions, we find that: a) language-specific code tokenizers outperform English tokenizers, and b) code entities at the method- and class-level can comfortably fit the models with the largest of input windows, but fitting larger contexts beyond class-level may still not be practical.

We find that a model that efficiently encodes its code input using a code-specific tokenizer, is able to encode the same data in less space. This leads to a greater amount of context-fitting. Therefore, we need to encourage researchers and model architects to adopt such changes, instead of relying on sub-tokenizations from tokenizers trained on English text. 

It's worth noting that classical Transformer models exhibit a quadratic complexity in terms of the input size due to the attention layers. This contributes to their issues in scaling beyond a threshold limit. Thus, reasoning at the scale of packages or projects would require a rethink of the architecture, such as using a Transformer variant that better handles longer sequences such as a Reformer \cite{kitaev2020reformer}, or another \textit{efficient} Transformer \cite{tay2020efficient} which exhibits lower complexities as input size increases. Whether this is sufficient is uncertain: \textit{efficient} transformers can struggle with very long sequences, as exhibited in specialized benchmarks \cite{tay2020long}.

While we focused specifically on Transformers as they have a fixed context size window, other models will also be challenged by large input sizes. The \texttt{ASTs} and graph representations of classes, packages, and projects will also have scaling issues as the number of nodes to consider will grow very quickly. Furthermore, Graph Neural Networks can also struggle with long-distance relationships in the graph \cite{alon2020bottleneck}. Clearly, significant work is needed to find architectures that can fit contexts at the project-level, especially if the model size is to be kept small enough to be manageable.

On the other hand, we see promise in an approach that is able to select the input relevant to the task. Of note, recent work has started to go in this direction for code summarization, both at the file level \cite{clement2021long} and multiple files \cite{bansal2021project}. Significant work lies ahead in devising techniques that truly take into account a larger global context, thus addressing the \textit{``Out-Of-Window''} \texttt{(OOW)} problem; at a minimum, \dataset provides the data at scale, and tools to investigate this.

\section{Limitations}
\label{sec:limitations}

\dataset is the only effort we are aware of in gathering enough data that is preprocessed sufficiently to enable empirical research of machine learning models that can reason on a more global context than the file or method level. Nevertheless, it has several limitations. Some of these issues are inherited from our use of 50K-C, while others are due to limitations in our pre-processing; while the former will be hard to overcome (barring extensive additional data collection), the latter could be mitigated by further processing from our side.

\subsection{Limitations stemming from the use of 50K-C}

\paragraph{Monolingual.} \dataset is comprised of projects in the Java programming language only. This poses issues as to whether models that work well for Java would also work well for other languages. The reason for this limitation is twofold: 1) adding other languages at a similar scale would drastically increase the already extremely significant time we invested in pre-processing data, and 2) restricting to one language frees us from tooling issues: we don't need to settle on a ``common denominator'' in tool support (e.g., Infer supports few programming languages, and many of  its analyses are limited to a single programming language).

\paragraph{Monoversion.} \dataset is comprised of snapshots of projects, rather than multiple project versions. This prevents us from using it for tasks that would rely on multiple versions, or commit data, such as some program repair tasks. On the other hand, this frees us from issues related to the evolution of software systems, such as performing origin analysis \cite{godfrey2005using}, which is essential as refactorings are very common in software evolution, and can lead to discontinuities in the history of entities, particularly for the most changed ones \cite{hora2018assessing}. Omitting versions also considerably reduces the size of the dataset, which is already rather large as it is.

\paragraph{Static data only.} While the projects included in \texttt{50K-C} were selected because they could be compiled, \texttt{50K-C} provide no guarantees that they can be run. Indeed, it is hard to know if a project can run, even if it can be compiled. In case it can run, the project likely expects some input of some sort. This leaves running test cases as the only option to reliably gather runtime data. In our previous work in Smalltalk, where we performed an empirical study of 1,000 Smalltalk projects, we could run tests for only 16\% of them \cite{callau2014use}. Thus, \dataset makes no attempt at gathering properties that comes from dynamic analysis tools at this time. In the future, \dataset's property mechanism could be used to document whether a project has runnable test cases, as a first step towards gathering runtime information. We could also expand the dataset with the 76 projects coming from XCorpus, which were selected because they are runnable \cite{dietrich2017xcorpus}.

\subsection{Limitations stemming from our pre-processing}

\paragraph{Incomplete compilation.} While the projects in \texttt{50K-C} were selected because they were successfully compiled, we were not able to successfully recompile all of them. Roughly 18\% of the largest projects could not be compiled; this number trends down for smaller projects. We are not always sure of the reasons for this, although we suspect that issues related to dependencies might come into play. This could add a bias to our data, in case the projects that we are unable to compile are markedly different from the ones that we could compile. Nevertheless, all of the meta-data, call-graphs, and almost all of the properties and representations could be generated even for uncompiled projects.

\paragraph{Imprecisions in call graphs.} The call graph extraction tool that we use has some limitations that we inherit. In particular, handling methods called via reflection is a known problem for static analysis \cite{bodden2011taming}; the call graph extraction tool does not handle these cases. A second issue is related to polymorphism, where it is impossible to know, in the absence of runtime information, which of the implementations can be called. In this case, our call graph has an edge to the most generic method declaration. 

\paragraph{Inner classes.} Our handling of inner classes is limited. Since inner classes are contained in methods, the models can have access to their definitions. However, we do not assign \texttt{UUIDs} to them or to the methods defined in them, as this would significantly increase the complexity of our model (in terms of levels of nesting in the hierarchy), while these cases are overall rare. Additional pre-processing could handle these cases, but we do not expect this to become necessary.

\paragraph{Class-level data.} Since most machine learning models of code take method-level samples as input, we work with this representation in our experiments, although we include larger contexts. As a consequence, our modeling of classes and packages is limited in this paper. While information about, for instance, the class attributes is not explicitly modeled in our work, it is easily accessible in the file-level feature graph representations, so that models that wish to use this information can access it.

\paragraph{Incomplete preprocessing.} At the time of writing, not all the representation data is present for all the projects, due to the very computationally expensive processing that is needed. We started with the largest projects, and worked our way down to the smaller ones. All of the metadata is present for all of the projects. However, some of the smaller projects (the ones with less than 20 classes) will have their representations computed and added to \dataset in the coming weeks. A second category of incomplete processing is that some tools will occasionally fail on some very specific input (e.g., the parser used by an analysis tool may handle some edge cases differently than the official parser).

\section{Conclusion}
\label{sec:conclusion}
In this article, we presented \dataset, a dataset and workbench to support research in the design and evaluation of source code machine learning models. Seen as a dataset, \dataset is built upon the \texttt{50K-C} dataset of 50,000 compilable Java projects, which we extend in several ways. We add multiple source code representations at the method level, to allow researchers to experiment on the effectiveness of these, and their variations. We add a project-level call graph, so the researchers can experiment with models that consider multiple methods, rather than a single method or a single file. Finally, we add multiple source code properties, obtained by running source code static analyzers---ranging from basic metrics to advanced analyses characteristics based on abstract interpretation.

\dataset \textit{Workbench}, its toolchain and corresponding \texttt{APIs}, help achieve a variety of objectives. \dataset can extend itself with new properties and representations. It can be used to define machine learning tasks, using the properties and the representations themselves as basis for prediction tasks. The properties defined in \dataset can be used to get insight into the performance of tasks and pinpoint possible sources of bias. Finally, \dataset provides all the tools to experiment with new representations that combine the existing ones, allowing the definition of models that can learn from larger contexts than a single method snippet.

Alongside, we have provided examples of usage of \dataset. We have shown how \dataset can be used to define a metric prediction and a method call completion task. We have also shown how \dataset can be used for empirical studies. In particular, we investigated how the performance of our code completion task was impacted by the type of identifier to predict, showing that models performed much better on API method calls than on method calls defined in the project, indicating the need for models that take into account the project's context. Finally, we have shown that taking into account this global context will be challenging, by studying its size. While state-of-the-art transformer models such as \texttt{CodeBERT} can fit most methods in the dataset, fitting package-level or higher context is much more challenging, even for the largest models such as OpenAI's \texttt{Codex} model. This indicates that significant effort lies ahead in defining models able to process this amount of data, a task that we hope \dataset will support the community in achieving.

\section{Declarations}
\subsection{Funding, Competing Interests, and Conflict of Interest}
\begin{itemize}
    \item[$\circ$] The authors have no relevant financial or non-financial interests to disclose.
    \item[$\circ$] The authors have no financial or proprietary interests in any material discussed in this article. Neither are there any usage of third-party artifacts.
    \item[$\circ$] This work was partially funded by the \texttt{IDEALS} and \texttt{ADVERB} projects of the Free
University of Bozen-Bolzano. The authors have no other competing interests to declare that are relevant to the content of this article.    
    \item[$\circ$] The authors declare that they have no conflict of interest.
\end{itemize}

\section{Data Availability Statements}
The particular datasets generated during and/or analysed during the current study (specifically, in Section~\ref{sec:non-localness}) are made available in the following repository: \\ \url{https://github.com/giganticode/jemma/tree/master/jemma/paper/}.

The official \dataset repository provides documentation, all links to datasets, and any other relevant information: \url{https://github.com/giganticode/jemma/}

\bibliographystyle{spbasic}
\bibliography{references}

\newpage
\begin{appendices}
\renewcommand\thefigure{\thesection.\arabic{figure}}  

\section{}
\label{appendix:scripts}
\setcounter{figure}{0}
\setcounter{table}{0}
\renewcommand{\thetable}{A\arabic{table}}

\subsection{\textbf{Building larger contexts}}
\label{appendix:a1}

Figure A.1 is an illustration of how larger contexts can be gradually created from callers/callees over 1-hop, 2-hop, and greater neighborhood of method calls. For a given method (000), its direct callee methods (001 and 002) provide additional context information from its immediate neighborhood (1-hop or level-1 neighborhood). Further callees (003, 005, 007) of the direct callee methods provide indirect context information from a further neighborhood (2-hop or level-2 neighborhood) and so on.

Based on the call graph, a given method can have multiple sources of context information from different neighborhoods (level-n) of callers and callers. Then, depending on the task, specific parts of the multiple sources of context information may be added to the input to establish the context information (e.g., method names, call statements, identifiers, entire method text). For our experiment in Section~\ref{sec:non-localness} we have considered the context information from a method's 1-hop neighborhood, considering all possible callee method names.

\begin{figure}[H]
  \frame{\includegraphics[width=100mm]{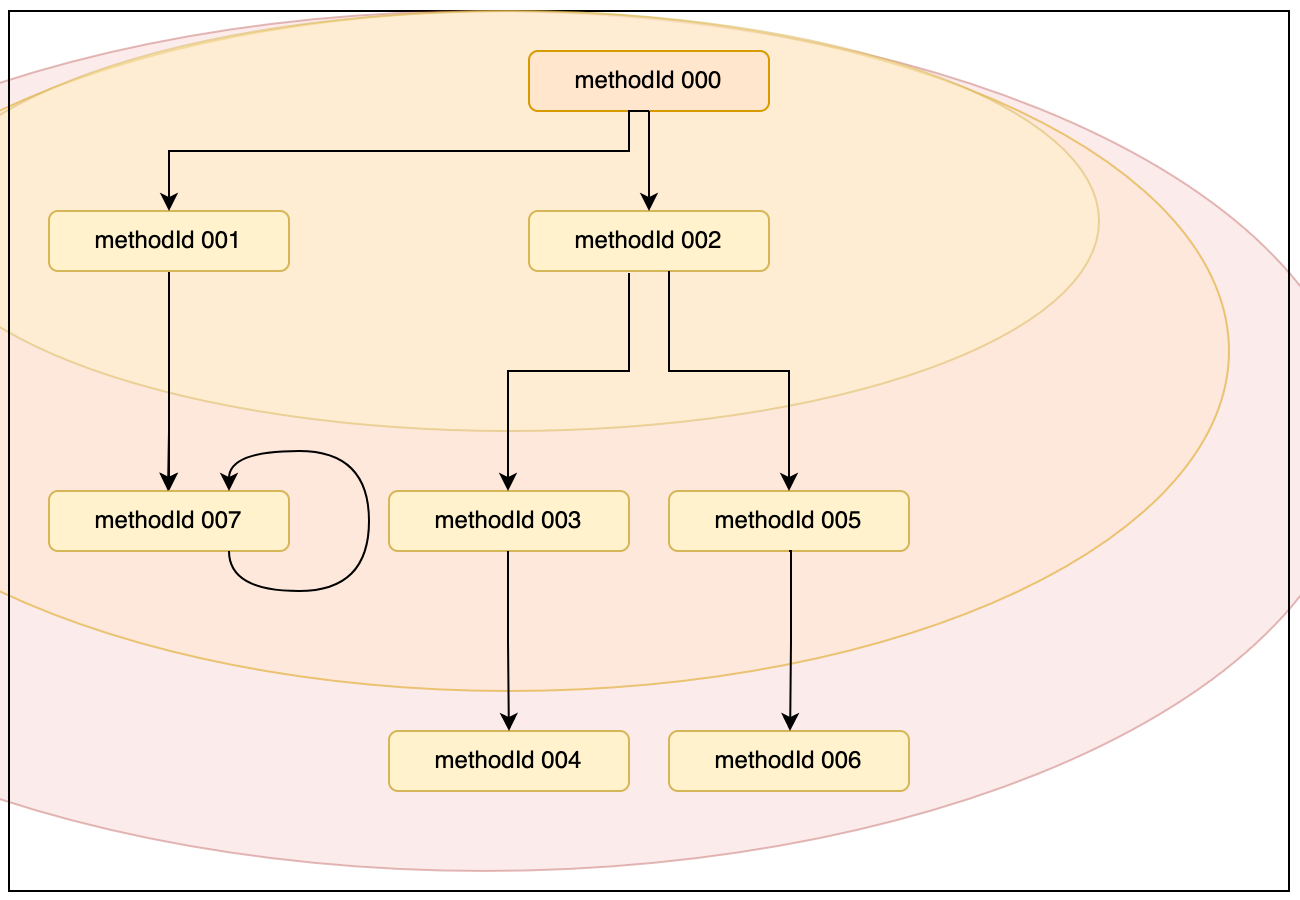}}
\captionsetup{labelformat=empty}
\caption{Step-wise inclusion of greater context information, based on \\
caller-callee relationships, from 1-hop, 2-hop, and 3-hop neighborhoods.}
\label{fig:creating-global-contexts}       
\end{figure}

\noindent
This example is representative of one of the ways in which how larger contexts may be constructed
Other techniques might consider including context from sibling methods in the same class, or from methods based on data-flow, control-flow, or dependency graphs. The use of context information while modeling source code is still nascent, therefore, further research on how to effectively build and use project-wide contexts is an open problem.

\newpage
\subsection{\textbf{Model accuracies by System Sizes}}
\label{appendix:a2}

\begin{table}
\label{tab:completion:sizesxxmm}
\caption{{Method call completion (by size) without vs. with context, and \% improvement.}  \\ 
        Scores for: \texttt{A - {BERT}, B - {CodeBERTa}, C - {CodeBERT}, D - {GraphCodeBERT}} } 
    \resizebox{\textwidth}{!}{%
    \begin{tabular}{|l|
    c|>{\columncolor[gray]{0.9}}c|c|
    c|>{\columncolor[gray]{0.9}}c|c|
    c|>{\columncolor[gray]{0.9}}c|c|
    c|>{\columncolor[gray]{0.9}}c|c|}
        \hline
        \multicolumn{1}{|c|}{\multirow{2}{*}{}}                                        
            & \multicolumn{3}{c|}{\textit{01-20}}                         
            & \multicolumn{3}{c|}{\textit{21-50}} 
            & \multicolumn{3}{c|}{\textit{51-100}}
            & \multicolumn{3}{c|}{\textit{\textgreater 100}} \\ 
         
        \hline %
        \multicolumn{1}{|c|}{}                 
            & \texttt{n/c} & \texttt{c}  & $\pm$
            & \texttt{n/c} & \texttt{c}  & $\pm$
            & \texttt{n/c} & \texttt{c}  & $\pm$
            & \texttt{n/c} & \texttt{c}  & $\pm$
            \\ \hline

\multicolumn{1}{|l|}{\texttt{A}}
 &  0.139  &  0.175  &  \textcolor{teal}{ 26\%}   	 &  0.155  &  0.203  &  \textcolor{teal}{ 31\%}   	 &  0.170  &  0.220  &  \textcolor{teal}{ 29\%}   	 &  0.194  &  0.237  &  \textcolor{teal}{ 22\%}   	 \\ \hline
\multicolumn{1}{|l|}{\texttt{B}}
 &  0.290  &  0.360  &  \textcolor{teal}{ 24\%}   	 &  0.315  &  0.419  &  \textcolor{teal}{ 33\%}   	 &  0.340  &  0.449  &  \textcolor{teal}{ 32\%}   	 &  0.331  &  0.426  &  \textcolor{teal}{ 29\%}   	 \\ \hline
\multicolumn{1}{|l|}{\texttt{C}}
 &  0.177  &  0.200  &  \textcolor{teal}{ 13\%}   	 &  0.211  &  0.247  &  \textcolor{teal}{ 17\%}   	 &  0.219  &  0.252  &  \textcolor{teal}{ 15\%}   	 &  0.245  &  0.277  &  \textcolor{teal}{ 13\%}   	 \\ \hline
\multicolumn{1}{|l|}{\texttt{D}}
 &  0.175  &  0.201  &  \textcolor{teal}{ 15\%}   	 &  0.212  &  0.247  &  \textcolor{teal}{ 17\%}   	 &  0.225  &  0.253  &  \textcolor{teal}{ 12\%}   	 &  0.249  &  0.278  &  \textcolor{teal}{ 12\%}   	 \\ \hline
 
    \end{tabular}}
\end{table}

Table A1 shows the accuracies for the method-call completion task, with and without additional context, split across project sizes. Once again, we observe that including additional non-local context improves the accuracy for all of the models. The \texttt{CodeBERTa} model posts the highest accuracies across all project sizes. Interestingly, it is also the model with the most efficient source code tokenizer compared to the others. This matches the observations in Section~\ref{sec:non-localness} where the results are split across call-types. 

To observe the call-type accuracy comparison (with and without context-information) for each project size, we have included Fig~\ref{fig:across_sizes}. The rows represent the models. The columns represent the project sizes. The red bars record the accuracies without context, while the green bars record the accuracies with added context --- for each call type: class, package, project, api. We can clearly see that for each model, the accuracies are similar across project sizes for the different call types, especially for package, project, api calls. Most impotantly, all models appear to improve their scores with added context across project-sizes and call-types.

\begin{figure}[H]
  \frame{\includegraphics[width=\textwidth, height=190pt]{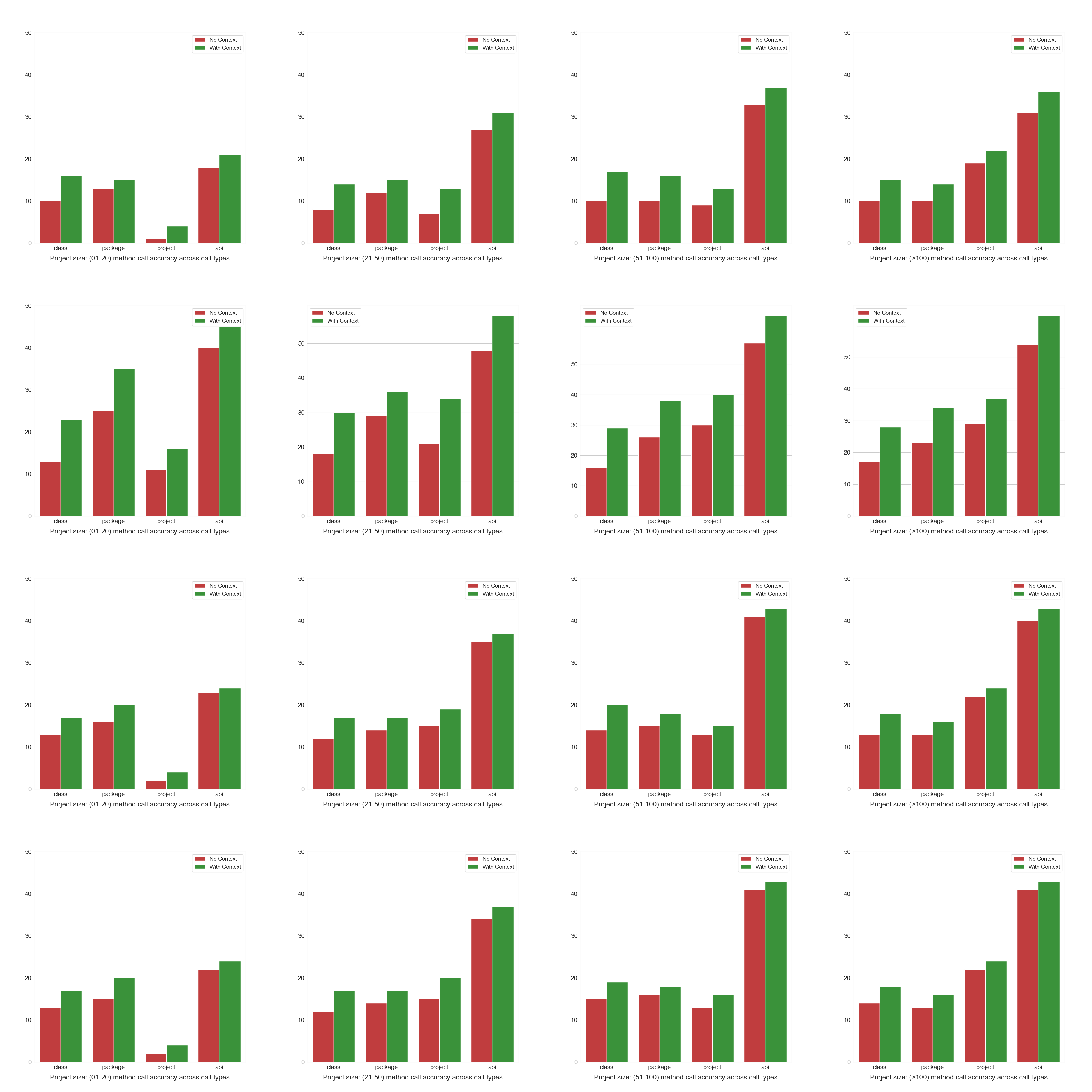}}
\captionsetup{labelformat=empty}
\caption{Call-type comparison for each project size.}
\label{fig:across_sizes}       
\end{figure}

\end{appendices}

\end{document}